\documentclass[longauth]{aa}  

\usepackage{graphicx}
\usepackage{txfonts}
\usepackage[hidelinks,colorlinks=true,linkcolor=blue,citecolor=blue]{hyperref}
\usepackage{color}
\usepackage[dvipsnames]{xcolor}
\usepackage{siunitx}

\newcommand{\hi}{H{\sc i}}

\begin{document} 

\title{The BINGO Project VII:} 
\subtitle{Cosmological forecasts from 21cm intensity mapping}

\author{Andre A. Costa\inst{1}\thanks{andrecosta@yzu.edu.cn}, 
Ricardo G. Landim\inst{2}, 
Camila P. Novaes\inst{3}, 
Linfeng Xiao\inst{4}, 
Elisa G. M. Ferreira\inst{5,6}, 
Filipe B. Abdalla\inst{3,5,7,8}, 
Bin Wang\inst{1,4}, 
Elcio Abdalla\inst{5},
Richard A. Battye\inst{9},
Alessandro Marins\inst{5},
Carlos A. Wuensche\inst{3},
Luciano Barosi\inst{10},
Francisco A. Brito\inst{10,11},
Amilcar R. Queiroz\inst{10}, 
Thyrso Villela\inst{3,12,13},
Karin S. F. Fornazier\inst{5},
Vincenzo Liccardo\inst{3},
Larissa Santos\inst{1,4},
Marcelo V. dos Santos\inst{10},
Jiajun Zhang\inst{14}
}

\institute{
Center for Gravitation and Cosmology, College of Physical Science and Technology, Yangzhou University, Yangzhou 225009, China
\and 
Technische Universit\"at M\"unchen, Physik-Department T70, James-Franck-Stra$\beta$e 1, 85748 Garching, Germany
\and 
Divis\~{a}o de Astrof\'{\i}sica, Instituto Nacional de Pesquisas Espaciais (INPE), S\~{a}o Jose dos Campos - SP, Brazil
\and 
School of Aeronautics and Astronautics, Shanghai Jiao Tong University, Shanghai 200240, China
\and 
Instituto de F\'isica, Universidade de S\~ao Paulo, C.P. 66.318, CEP 05315-970, S\~ao Paulo, Brazil
\and 
 Max-Planck-Institut f{\"u}r Astrophysik, Karl-Schwarzschild Str. 1, 85741 Garching, Germany
\and 
Department of Physics and Astronomy, University College London, Gower Street, London, WC1E 6BT, UK,
\and 
Department of Physics and Electronics, Rhodes University, PO Box 94, Grahamstown, 6140, South Africa,
\and 
Jodrell Bank Centre for Astrophysics, Department of Physics and Astronomy, The University of Manchester, Oxford Road, Manchester, M13 9PL, U.K. 
\and 
Unidade Acad\^emica de F\'{i}sica, Universidade Federal de Campina Grande, R. Apr\'{i}gio Veloso, 58429-900 - Bodocong\'o, Campina Grande - PB, Brazil
\and 
Departamento de F\'{i}sica, Universidade Federal da Para\'{i}ba, Caixa Postal 5008, 58051-970 Jo\~{a}o Pessoa, Para\'{i}ba, Brazil
\and 
Instituto de F\'{i}sica, Universidade de Bras\'{i}lia, Bras\'{i}lia, DF, Brazil
\and 
Centro de Gest\~ao e Estudos Estrat\'egicos - CGEE,
SCS Quadra 9, Lote C, Torre C S/N Salas 401 - 405, 70308-200 - Bras\'ilia, DF, Brazil
\and 
Center for Theoretical Physics of the Universe, Institute for Basic Science (IBS), Daejeon 34126, Korea \\
             }
             
   \authorrunning{A. A. Costa et al.}
   \titlerunning{The BINGO Project VII: Cosmological forecasts from 21cm intensity mapping}

   \date{Received Month Day, Year; accepted Month Day, Year}

 
  \abstract
   {The 21cm line of neutral hydrogen (\hi) opens a new avenue in our exploration of the  structure and evolution of the Universe. It provides complementary data to the current large-scale structure (LSS) observations with different systematics, 
   and thus it will be used to improve our understanding of the Lambda Cold Dark Matter ($\Lambda$CDM) model. This will ultimately constrain our cosmological models, attack unresolved tensions, and test our cosmological paradigm. Among several radio cosmological surveys designed to measure this line, BINGO is a single-dish  telescope mainly designed to detect baryon acoustic oscillations (BAOs) at low redshifts ($0.127 < z < 0.449$).}
   {{Our goal is to assess  the fiducial BINGO setup and its capabilities of
constraining the cosmological parameters, and to analyze the effect of different
instrument configurations.} }
   {We  used the 21cm angular power spectra to extract cosmological information about the \hi~ signal and the Fisher matrix formalism to study BINGO's projected constraining power.}
   {We used the Phase 1 fiducial configuration of the BINGO telescope to perform our cosmological forecasts. In addition, we investigated the impact of several instrumental setups, taking into account some instrumental systematics, and different cosmological models. Combining BINGO with \textit{Planck} temperature and polarization data,
   the projected constraint improves from a $13\%$ and $25\%$ precision measurement at the 68\% confidence level with \textit{Planck} only to $1\%$ and $3\%$ for the Hubble constant and the dark energy (DE) equation of state (EoS), respectively, within the wCDM model. Assuming a Chevallier–Polarski–Linder (CPL) parameterization, the EoS parameters have standard deviations given by $\sigma_{w_0} = 0.30$ and $\sigma_{w_a} = 1.2$, which are improvements  on the order of $30\%$ with respect to \textit{Planck} alone. We  also compared BINGO's fiducial forecast with future SKA measurements and found that, although it will not provide competitive constraints on the DE EoS, significant information about \hi\, distribution can be acquired. We can access information about the \hi\, density and bias, obtaining $\sim 8.5\%$ and $\sim 6\%$ precision, respectively, assuming they vary with redshift at three independent bins.
   BINGO can also help constrain alternative models, such as interacting dark energy and modified gravity models, improving the cosmological constraints significantly. }
   {The fiducial BINGO configuration will be able to extract significant cosmological information from the \hi\, distribution and provide constraints competitive with current and future cosmological surveys. It will also help in understanding the \hi\,  physics and systematic effects.}

   \keywords{Cosmology -- Baryon Acoustic Oscillations --
                21cm Intensity Mapping
               }

   \maketitle
%

\section{Introduction}
\label{sec:intro}
Since the first direct measurement indicating the Universe undergoes an accelerated expansion phase from type Ia supernovae \citep[SNIa;][]{Perlmutter:1998np,Riess:1998cb}, several other observations have been accumulated. They  strengthen the evidence in favor of an accelerated phase in the Universe, whose standard driving force candidate is a cosmological constant, $\Lambda$.

In the current era of precision cosmology, measurements of the cosmic microwave background (CMB) by the \textit{Planck} satellite, among others, provide constraints on the parameters of the standard Lambda Cold Dark Matter ($\Lambda$CDM) model with high precision 
\citep{Aghanim:2018eyx}. Other probes provide additional information about the Universe's evolution and are essential to indicate whether alternatives to the $\Lambda$CDM model are more suitable to explain the current observations \citep{Abdalla:2020ypg}. In this direction, measurements of the 21cm line of neutral hydrogen (\hi) are expected to be one of the leading cosmological probes in the next years, opening a new avenue to survey the large-scale structure (LSS) of our Universe \citep{Pritchard:2011xb}.

\hi\, is a biased tracer of the galaxy distribution. Although the \hi\, distribution can be resolved as in an optical galaxy survey \citep{Bacon:2018dui}, its radio line requires a very large collecting area to obtain the necessary sensitivity for its detection. On the other hand, we can also map the total \hi\, intensity on large angular scales with much smaller instruments \citep{Battye:2012tg}. Intensity mapping (IM) measurements allow us to probe large volumes of the Universe in a much shorter amount of time compared with optical surveys where galaxies have to be resolved \citep{Battye:2004re,Chang:2007xk,Loeb:2008hg,Sethi:2005gv,Visbal:2008rg}. The technique is similar to measuring the CMB radiation, but with added redshift information.
It is especially suited to measuring baryon acoustic oscillations (BAOs) in the post-reionization era, which are imprinted at large cosmological scales. Due to this combination of large volumes and extra evolutionary information, \hi\,  IM is a powerful and competitive probe in cosmology.

The first detection of \hi\, in a cosmological survey became a proof of concept that the IM technique can indeed be used to probe the LSS of our Universe \citep{Kerp:2011xm, Chang:2010jp,Switzer:2013ewa, Masui:2012zc}. Even though these measurements were limited in area and sensitivity to extract cosmological information,
they highlighted the challenges of such detections: systematics effects present in the observed \hi\,  data and astrophysical foregrounds.

Systematic effects can come from the unknown cosmological evolution of both the \hi\, average density and bias. They can also be due to the $1/f$ noise present in the experiment \citep[see, e.g., ][]{maino99,Seiffert:2002gh,Meinhold09}, standing waves  \citep{RohlfsWilson:2004}, and contamination from radio frequency interference (RFI) at the site \citep{Harper:2018b}. The presence of foregrounds is still one of the main challenges for the \hi\, signal detection \citep{Battye:2012tg,Bigot-Sazy:2015jaa}. They originate from galactic and extra-galactic sources and can be orders of magnitude above the \hi\, signal. Our ability to remove foregrounds and properly understand systematic effects is crucial to adequately extracting the \hi\, signal that can be used for cosmological studies.

Several ongoing and upcoming telescopes will use the IM technique to measure BAOs from the 21cm line, such as the Canadian Hydrogen Intensity Mapping Experiment\footnote{\url{https://chime-experiment.ca/}}  \citep[CHIME;][]{Bandura:2014gwa}, Five-hundred-meter Aperture Spherical Radio Telescope\footnote{\url{http://fast.bao.ac.cn/en/FAST.html}} \citep[FAST;][]{Nan:2011um}, Square Kilometer Array\footnote{\url{https://www.skatelescope.org/}}  \citep[SKA;][]{Santos:2015gra}, Tianlai\footnote{\url{http://tianlai.bao.ac.cn/wiki/index.php/Main_Page}} \citep{Chen:2012xu}, and BAO from Integrated Neutral Gas Observations\footnote{\url{https://www.bingotelescope.org/en/}}  \citep[BINGO;][]{Battye:2012fd,Battye:2016qhf,Wuensche:2018,2020_project}. BINGO aims to measure the \hi\, IM at low-z precisely enough to constrain the cosmological parameters; it is  projected to be a Stage II\footnote{Classification according to the Dark Energy Task Force \citep{Albrecht:2006um}.} probe, competitive in the context of current and future cosmological surveys.

One of the first steps of any proposed experiment is to access its {capability of providing} useful data and its theoretical ability to constrain parameters. In this sense, it is necessary to forecast its ability and precision to extract valuable physical information and constrain various models. In this paper we forecast the potential of BINGO to constrain the 
cosmological parameters, assuming an ideal data output, and to help us understand the properties of DE, which is one of the main goals of this experiment \citep{2020_project}.

This is  paper VII in a series of papers describing the BINGO project. The theoretical and instrumental projects are in papers I and II  \citep{2020_project, 2020_instrument}, the optical design in paper III \citep{2020_optical_design}, the mission simulation in paper IV \citep{2020_sky_simulation}, further steps in component separation and bispectrum analysis in paper V
\citep{2020_component_separation}, and a mock is described in paper VI \citep{2020_mock_simulations}.
 
This paper is organized as follows. Section~\ref{sec:21cm_Cl} presents the theoretical 21cm angular power spectra, which will be used to constrain our cosmological models. In Sect.~\ref{sec:fisher} we introduce the Fisher matrix formalism that is considered in our forecasts. Our results are described in Sect.~\ref{sec:results} for several different experiment configurations and cosmological models. Finally, we summarize our conclusions in Sect.~\ref{sec:conclusions}. In  Appendix~\ref{sec:appendix} we compare two independent 21cm angular power spectrum codes developed by members of our collaboration, one of which is  used throughout the present analysis.

\section{21cm angular power spectra}
\label{sec:21cm_Cl}
The 21cm line of \hi\, originates from the hyperfine structure of the hydrogen atom. Some astrophysical mechanisms can lead to a change in the \hi\,  state and produce such a line, which is observed (redshifted) on Earth \citep[for a review of 21cm Cosmology, see, e.g.,][]{Furlanetto:2006jb,Pritchard:2011xb}. We relate the photon distribution coming from these sources by the 21cm brightness temperature, $T_b$, which at the background level is given by \citep{Hall:2012wd}
\begin{align}\label{eq:Tb_mean}
    \bar{T}_b(z) & = \frac{3(h_pc)^3\bar{n}_{\textrm{\hi\,}} A_{10}}{32 \pi k_B E_{21}^2 (1+z) H(z)} \nonumber \\
    & = 188 \,h\,\Omega_{\textrm{\hi\,}}(z)\frac{(1 + z)^2}{E(z)} \, \text{mK} \,.
\end{align}
Here $A_{10}$ is the spontaneous emission coefficient, $\bar{n}_{\textrm{\hi\,}}$ is the rest-frame average number density of \hi\, atoms at redshift $z$, $E_{21}=5.88\times 10^{-6}$ eV is the difference between the two energy levels associated with the \hi\, hyperfine splitting, $E(z) = H(z)/H_0$ is the normalized Hubble parameter, where the Hubble constant is defined by $H_0 = 100 h $ km s$^{-1}$Mpc$^{-1}$, $\Omega_{\textrm{\hi\,}}(z)$ describes the \hi\, density parameter in units of the current critical density, and, finally, $c$, $h_p$, and $k_B$ are the light speed, the Planck constant, and the Boltzmann constant, respectively.

At large scales we can treat inhomogeneities and anisotropies assuming small perturbations around the homogeneous and isotropic background. Matter density perturbations will feed the gravitational potentials, which will in turn modify the density distribution. Assuming the conformal Newtonian gauge, the metric takes the form
\begin{equation}\label{eq:ds}
    ds^2 = a^2(\eta)\Big[(1+2\Psi(\eta,\Vec{x}))d\eta^2-(1-2\Phi(\eta,\Vec{x}))d\Vec{x}^2\Big]\,,
\end{equation}
where $\eta$ is the conformal time, $a(\eta)$ is the scale factor, and $\Psi$ and $\Phi$ are the space-time gravitational potentials. Assuming linear order perturbations, the fractional brightness temperature perturbation in the $\hat{\mathbf{n}}$ direction, at redshift $z$, is \citep{Hall:2012wd}
\begin{align}\label{eq:frac_Tb}
    \Delta_{T_b} (z,\hat{\mathbf{n}})& = \delta_{\textrm{\hi\,}} - \frac{1}{\mathcal{H}} \, \hat{\mathbf{n}} \cdot \left( \hat{\mathbf{n}} \cdot \nabla \mathbf{v} \right) + \left( \frac{d \ln (a^3 \bar{n}_{\textrm{\hi\,}})}{d \eta} - \frac{\dot{\mathcal{H}}}{\mathcal{H}} -2 \mathcal{H} \right) \delta \eta \nonumber \\
& + \frac{1}{\mathcal{H}} \dot{\Phi}+\Psi \,,
\end{align}
where $\delta_{\textrm{\hi\,}}$ is the \hi\, density perturbation, $\mathbf{v}$ is its peculiar velocity, and $\mathcal{H}$ is the Hubble parameter in conformal time. This expression takes into account all relativistic and line-of-sight components at post-reionization era\footnote{At low redshift, the spin temperature (which relates the population in the excited state of the 21cm line to the ground state) is much higher than the CMB temperature. Therefore, the results are independent of the spin temperature, and the stimulated emission and absorption can be neglected.}, assuming that the comoving number density of \hi\, is conserved at low redshifts and using the Euler equation $\dot{\mathbf{v}} + \mathcal{H}\mathbf{v} + \nabla\Psi = 0$.

As described in \cite{Hall:2012wd}, the terms in Eq.~(\ref{eq:frac_Tb}) have a simple physical explanation. The first term corresponds to the \hi\, density perturbation, which we relate to the underlying matter distribution through some bias.\footnote{Following \cite{Hall:2012wd}, we calculate the \hi\, density perturbation in the comoving gauge as $\delta_{\textrm{\hi\,}} = b_{\textrm{\hi\,}}\delta_m^{\textrm{syn}} + \left(\frac{d\ln{(a^3\bar{n}_{\textrm{\hi\,}})}}{d\eta} - 3\mathcal{H}\right)\frac{v}{k}$, where $b_{\textrm{\hi\,}}$ is the \hi\, bias, $\delta_m^{\textrm{syn}}$ is the synchronous-gauge matter overdensity, and $v$ is the Newtonian-gauge matter velocity.} The second term is the redshift-space distortion (RSD) component. The third term originates from the zero-order brightness temperature calculated at the perturbed time of the observed redshift. The fourth term comes from the part of the integrated Sachs-Wolfe (ISW) effect that is not canceled by the Euler equation. Finally, the last component arises from increments in redshift from radial distances in the gas frame.

Due to the full sky characteristic of 21cm radio surveys, it is natural to decompose the brightness temperature in spherical harmonics. Therefore, for a fixed redshift, we have
\begin{equation}
    \Delta_{T_b} (z,\hat{\mathbf{n}})= \sum_{\ell m} \Delta_{T_b, \ell m} (z) Y_{\ell m}(\hat{\mathbf{n}})\,.
\end{equation}
If we express these perturbations in terms of their Fourier transform, we obtain
\begin{align}
\Delta_{T_{b,\ell}}(\mathbf{k},z) &=  \delta_{\textrm{\hi\,}}j_\ell(k\chi) + \frac{kv}{\mathcal{H}}j_\ell''(k\chi) + \left(\frac{1}{\mathcal{H}}\dot{\Phi}+\Psi\right)j_\ell(k\chi) \nonumber \\
& - \left[\frac{1}{\mathcal{H}}\frac{d\ln(a^3\bar{n}_{\textrm{\hi\,}})}{d\eta} - \frac{\dot{\mathcal{H}}}{\mathcal{H}^2}-2\right]\Bigg[\Psi j_\ell(k\chi) + vj_\ell'(k\chi) \nonumber\\&+ \int_{0}^{\chi}(\dot{\Psi} + \dot{\Phi})j_\ell(k\chi ')d\chi '\Bigg] \,,
\end{align}
where $j_\ell (k\chi)$ are spherical Bessel functions, which depend on the wave number $k$ and the comoving distance $\chi$, and their primes denote derivatives with respect to the argument.

In order to extract information about the \hi\, distribution, we need to access the statistical properties of the 21cm brightness temperature signal. The first statistical moment is the average, which corresponds to the background value in Eq.~(\ref{eq:Tb_mean}). Assuming linear perturbations the second object, the one-point correlation function, should be zero. Therefore, the next term is the two-point correlation function or its Fourier transform, the power spectrum. 
This is related to the angular cross-spectra by
\begin{equation}\label{eq:cl}
C_\ell (z_i, z_j) = 4\pi \int d\ln{k} \mathcal{P_R}(k)\Delta_{T_{b,\ell}}^W(\mathbf{k}, z_i)\Delta_{T_{b,\ell}}^{W'}(\mathbf{k}, z_j)\, ,
\end{equation}
where $\mathcal{P_R}(k)$ corresponds to the dimensionless power spectrum of the primordial curvature perturbation $\mathcal{R}$, and
\begin{equation}\label{eq:DW}
\Delta_{T_{b,\ell}}^W(\mathbf{k},z) = \int_{0}^{\infty}dz\bar{T}_b(z)W(z)\Delta_{T_{b,\ell}}(\mathbf{k},z)
\end{equation}
sums up all the contributions to the signal in the redshift bin defined by the normalized window function $W(z)$.

\begin{figure}[h!]
\centering
\includegraphics[scale=0.60]{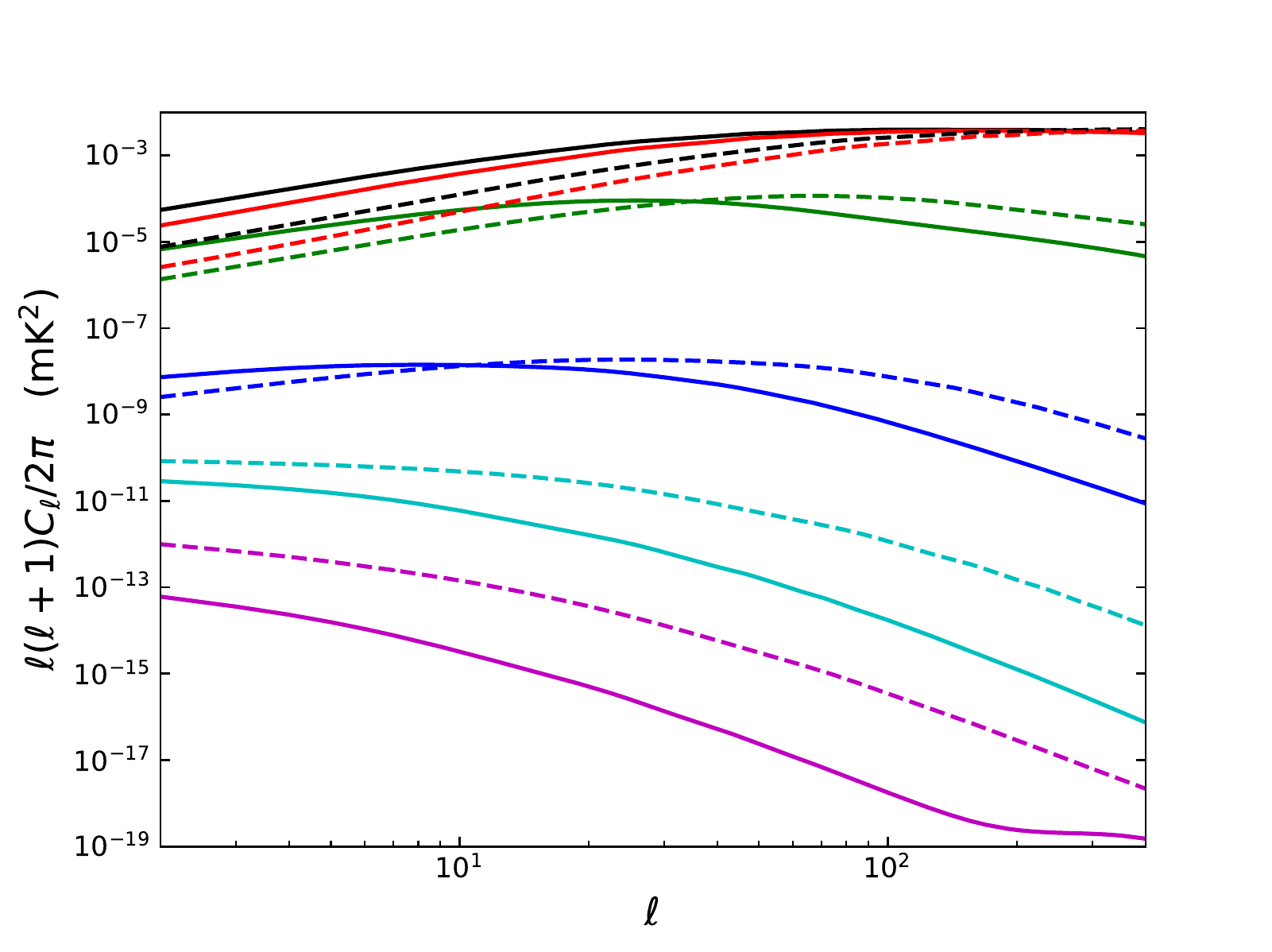}
\caption{\label{Cl_total} Brightness temperature perturbation power spectrum at $z = 0.13$ (solid lines) and $z = 0.45$ (dashed lines) with a 9.33 MHz bandwidth. The auto-spectra of the full signal (black) and of each individual term are shown, generically grouped as Newtonian-gauge density (red), redshift-space distortions (green), velocity term (blue), all potential terms evaluated at the source position (cyan), and the ISW component (magenta). The power spectrum is dominated by the \hi\, overdensity at small scales and has a significant contribution from RSD at large scales.}
\end{figure}

Figure~\ref{Cl_total} shows all the independent contributions to the 21cm angular spectrum at the first and last redshift bins of BINGO with a bandwidth of 9.33 MHz. We can observe that as we go to higher redshifts the contributions from the ISW effect and potentials increase. This happens because in a $\Lambda$CDM cosmology the gravitational potentials decrease with the scale factor at late times and, as we go to higher redshifts, the ISW effect sums the contributions over a wider range. The other terms in the 21cm spectrum behave differently at large and small scales. At the first redshift bin of BINGO, they are higher at the largest scales and are gradually  surpassed by the spectra of the last bin at small scales.

\section{The Fisher matrix}
\label{sec:fisher}
We  forecast the constraints from the upcoming 21cm IM BINGO telescope using the Fisher matrix formalism. The Fisher matrix for the parameters $\theta_i$ of a model $\mathcal{M}$ is defined as the ensemble average of the Hessian matrix of the log-likelihood. Assuming Gaussian probability distribution
with zero mean and covariance {\bf C}, the Fisher matrix can be calculated as \citep{Tegmark:1996bz} 
\begin{align}
F_{ij} \equiv \left\langle-\frac{\partial\ln \mathcal{L}}{\partial\theta_i \, \partial\theta_j}\right\rangle = \frac{1}{2}\mathrm{Tr}\left[\mathrm{\bf C}^{-1}\frac{\partial\mathrm{\bf C}}{\partial\theta_i}\mathrm{\bf C}^{-1}\frac{\partial\mathrm{\bf C}}{\partial\theta_j}\right] \;.
\end{align}
The covariance {\bf C} is the sum of the signal and noise spectra estimators. Considering only the thermal and shot noises, it can be written as
\begin{equation}
\label{eq:cov}
\mathrm{\bf C} = C_{\ell}(z_i, z_j) + C_{\ell}^{\rm{shot}}(z_i, z_j) +  N_{\ell}(z_i, z_j) \,.
\end{equation}
The inverse of the Fisher matrix gives the covariance among the parameters with diagonal elements corresponding to the $1\sigma$ marginalized constraints.

Analogously to what is done in the CMB case, we use the $a_{lm}$ values for the pixelization scheme, where
\begin{equation}
\left\langle a_{\ell m}(z_i)a_{\ell 'm'}^*(z_j) \right\rangle = \delta_{\ell \ell'}\delta_{mm'} C_\ell (z_i,z_j) \,.
\end{equation}
We note that for the CMB all the $a_{\ell m}$ values are evaluated at the same redshift, while the 21cm signal has information about a 3D volume that can be analyzed in a tomographic way. Therefore, to extend the covariance matrix to the case of a 21cm experiment, {we use our pixelization as} $a_{\ell m}(z)$. Then, the CMB diagonal matrix {is transformed into} a diagonal block matrix as
\begin{align}
\mathrm{\bf C} &=
\begin{bmatrix}
A_{\ell = 2} & 0 & ... & 0 \\
0 & A_3 & ... & 0 \\
\vdots & \vdots & ... & \vdots \\
0 & 0& ... & A_n
\end{bmatrix}
\,, \qquad \\ 
\text{where} \nonumber\\
A_\ell &= (2\ell + 1)
\begin{bmatrix}
C_\ell(z_1,z_1) & C_\ell(z_1,z_2) & ... & C_\ell(z_1,z_n) \\
C_\ell(z_2,z_1) & C_\ell(z_2,z_2) & ... & C_\ell(z_2,z_n) \\
\vdots & \vdots & ... & \vdots \\
C_\ell(z_n,z_1) & C_\ell(z_n,z_2)& ... & C_\ell(z_n,z_n)
\end{bmatrix}
\,.
\end{align}

\subsection{Model and parameters}

The $\Lambda$CDM model is the present cosmological paradigm; it is the simplest model in exquisite agreement with a wide range of cosmological data. In this model the Universe is composed of baryons, photons, neutrinos, DM, and DE, and the gravitational interaction among them is described by general relativity (GR). The Universe begins with an extremely dense and hot plasma. During an early exponential expansion phase, quantum fluctuations in the field driving inflation seeded inhomogeneities in the primordial plasma, providing the initial conditions to all the cosmological structures we observe today. CDM yields the necessary gravitational potentials to amplify the initial fluctuations and DE, assumed as a cosmological constant ($\Lambda$), is responsible for the late-time cosmic acceleration. These two components are responsible for about $95\%$ of the energy density budget today.

In this model we assume that the Universe is spatially flat and is governed by six cosmological parameters: $\Omega_b$ and $\Omega_c$, the baryon and DM density parameters, respectively, with $\Omega_i = \rho_i/\rho_c$, where $\rho_c$ is the critical density today; $h$, the Hubble constant parameter $H_0 = 100 h $ km s$^{-1}$Mpc$^{-1}$; the reionization optical depth, $\tau$; the amplitude and spectral index of primordial scalar density perturbations, $A_s$ and $n_s$, respectively.

The standard model assumes a DE given by a cosmological constant with EoS, $w = P/\rho = -1$. The simplest extension to this model consists in a dynamical DE with an EoS different from $-1$. In most of this paper, we  use the Chevallier–Polarski–Linder (CPL) parameterization \citep{Chevallier:2000qy,Linder:2002et} as our fiducial cosmological model, which allows us to study the constraints on the evolution of the DE EoS. The CPL parameterization is a $z$-dependent Ansatz for the EoS of DE given by
\begin{equation}\label{eq:eosDEcpl}
    w_{\text{CPL}}(z)=w_0+w_a\frac{z}{1+z} \quad \textrm{or} \quad w_{\text{CPL}}(a)=w_0+w_a(1-a)\,,
\end{equation}
where $w_0$ and $w_a$ are constants and $\Lambda$CDM is recovered when $w_0=-1$ and $w_a=0$. Therefore, we  vary the Fisher matrix with respect to the following set of cosmological parameters:
\begin{equation}\label{eq:ficudial_parameters}
    {\bf \theta} = \left\{\Omega_bh^2, \, \Omega_ch^2, \, h, \, \ln(10^{10}A_s), \, n_s, \, w_0, \, w_a, \, b_{\textrm{\hi\,}} \right\} \,.
\end{equation}
In addition, the 21cm spectra also depends on the \hi\, density parameter $\Omega_{\textrm{\hi\,}}$, which we fix to its fiducial value. Our ability to constrain this parameter with BINGO and its effect on the cosmological constraints is left to Sect.~\ref{sec:HI_density_bias}.

As our fiducial values we have chosen the best fit from \textit{Planck} 2018 \citep{Aghanim:2018eyx}, which we present in Table~\ref{tab:fiducial_param}. We have calculated the partial derivatives numerically with a step size of $\pm 0.5\% \times \theta_i$. The step size should not be too large, to avoid a miscalculated derivative, nor too small, introducing numerical noise. We  checked for the stability of our derivatives to this choice.

\begin{table}
\centering
\caption{\label{tab:fiducial_param} Fiducial values for the cosmological parameters in Eq. (\ref{eq:ficudial_parameters}) from \textit{Planck} 2018 \citep{Aghanim:2018eyx}. The last two (extra) parameters come from \hi\, physics, where we use the constant value for $\Omega_{\textrm{\hi\,}}$ measured by \cite{Switzer:2013ewa}. }
\begin{tabular}{lc}
\hline
Parameter & Fiducial value\\
\hline
$\Omega_b h^2$ & 0.022383 \\
$\Omega_c h^2$ & 0.12011 \\
$h$ & 0.6732\\
$n_s$ & 0.96605 \\
$A_s$ & $2.1\times 10^{-9}$ \\
$w_0$ & $-1$ \\
$w_a$ &  0 \\ 
$b_{\textrm{\hi\,}}$ & 1 \\ 
$\Omega_{\textrm{\hi\,}}$ & $6.2\times 10^{-4}$ \\
\hline
\end{tabular}

\end{table}

BINGO will help put constraints on the late-time Universe parameters. The combination with other probes can break degeneracies among several parameters and improve these constraints. In particular, {\it Planck} has provided precise CMB measurements, which gives tight constraints on the standard $\Lambda$CDM model. Therefore, we will combine our 21cm IM forecasts with a prior obtained from the {\it Planck} 2018 TT + TE + EE + lowE likelihood data \citep{Aghanim:2018eyx}. Using the publicly available code \texttt{CosmoMC} \citep{Lewis:2002ah},\footnote{\url{https://cosmologist.info/cosmomc/}} we {use the Markov chain Monte Carlo (MCMC) technique} to estimate the
covariance for the cosmological parameters from the {\it Planck} data. We then combine the {\it Planck} covariance with our 21cm IM Fisher matrices. It should be noted that, in general, the maximum likelihood from {\it Planck} will not coincide with our fiducial values. Here we assume that these constraints do not change significantly over the parameter space.

\subsection{Thermal noise}\label{sec:tehermal_noise}
The thermal noise describes the fundamental sensitivity  of a radio telescope. It corresponds to the voltages generated by thermal agitations in the resistive components of the antenna receiver. It appears as a uniform Gaussian distribution over the sky, with theoretical noise level per pixel calculated by the radiometer equation \citep{Wilson:2013}
\begin{equation}\label{eq:thermal_noise0}
    \sigma_T = \frac{T_{\rm{sys}}}{\sqrt{\Delta\nu \,t_{\rm{pix}}}} \,,
\end{equation}
where $T_{\rm{sys}}$ is the total system temperature (antenna plus sky temperatures), $\Delta\nu$ is the frequency channel width, and $t_{\rm{pix}}$ is the integration time per pixel, which is related to the total observational time $t_{\rm{obs}}$ by
\begin{equation}
    t_{\rm{pix}} = t_{\rm{obs}}n_{\rm{beam}}n_{f}\frac{\Omega_{\rm{pix}}}{\Omega_{\rm{sur}}} \,,
\end{equation}
where $n_f$ denotes the number of feed horns, $n_{\rm{beam}}$ is the number of {beams and polarizations} in each horn, and $\Omega_{\rm{pix}}$ and $\Omega_{\rm{sur}}$ describe the pixel and total survey area, respectively.

Assuming the thermal noise between different frequencies are uncorrelated, the noise covariance can be calculated as
\begin{equation}\label{eq:thermal_noise}
{N_\ell} = \sigma_T^2\Omega_{\rm{pix}} = \frac{T^2_{\rm{sys}}}{\Delta \nu\, t_{\rm{obs}}}\frac{\Omega_{\rm{sur}}}{2n_f} = \frac{T^2_{\rm{sys}}}{\Delta \nu\, t_{\rm{obs}}}\left(\frac{4\pi f_{\rm{sky}}}{2n_f}\right) \,,
\end{equation}
where $f_{\rm{sky}}$ is the surveyed fraction of the sky and we have assumed two polarizations are measured.

The telescope also has a maximum beam resolution that should be taken into account. This effect reduces the signal by a factor of $b_\ell^2$, which is given by \citep{Bull:2014rha}
\begin{equation}
    b_\ell (z_i) = \exp{\left(-\frac{1}{2}\ell^2\sigma_{b, \, i}^2\right)} \,,
\end{equation}
where $\sigma_{b, \, i} = \theta_B(z_i)/\sqrt{8\ln{2}}$ {with}
\begin{equation}
    \theta_B(z_i) = \theta_{\rm{FWHM}}(\nu_{\rm{center}})\frac{\nu_{\rm{center}}}{\nu_i}\,,
\end{equation}
where $\nu_{\rm{center}}$ is the middle frequency of the survey. Instead of reducing the signal power spectrum, we can think of the beam resolution as an increase in the noise by the inverse of $b_\ell^2$.
\begin{table}[h]
    \centering 
       \renewcommand*{\arraystretch}{1.4}
    \caption{ \label{tab:bingo_param}    Fiducial parameters of the BINGO telescope.}
    \begin{tabular}{@{}lc@{}}
        \hline 
        Parameters     &   BINGO    \\ 
       \hline
           Frequency range          & [980, 1260] MHz \\ 
      
           Redshift range           & [0.127, 0.449] \\
           
           Number of frequency channels\footnotemark       & 30 \\
      
           Number of feed horns     & 28  \\ 
     
           Sky coverage with Galactic mask\footnotemark  & 2900  deg$^2$ \\ 
    
           Observational time ($t_{\rm{obs}}$)       & 1 year \\
    
           System temperature ($T_{\rm{sys}}$)       & $70$ K    \\
     
           Beam resolution ($\theta_{\textrm{FWHM}}$)         & 40 arcmin \\
        \hline 
    \end{tabular}

\end{table}
\footnotetext[10]{According to BINGO's instrument paper \citep{2020_instrument}, the actual number of frequency channels is much larger than considered here. Our value takes into account a smoothing in the raw data for cosmological analysis.}
\footnotetext[11]{If we consider the sky coverage with the horns moving $\pm30 \, \textrm{cm}$ and a mask that removes about 20\% of the sky, the area effectively covered can be increased to about 4000 square degrees \citep{2020_optical_design}.}

\subsection{Shot noise}
\begin{figure}[h!]
\centering
\includegraphics[scale=0.55]{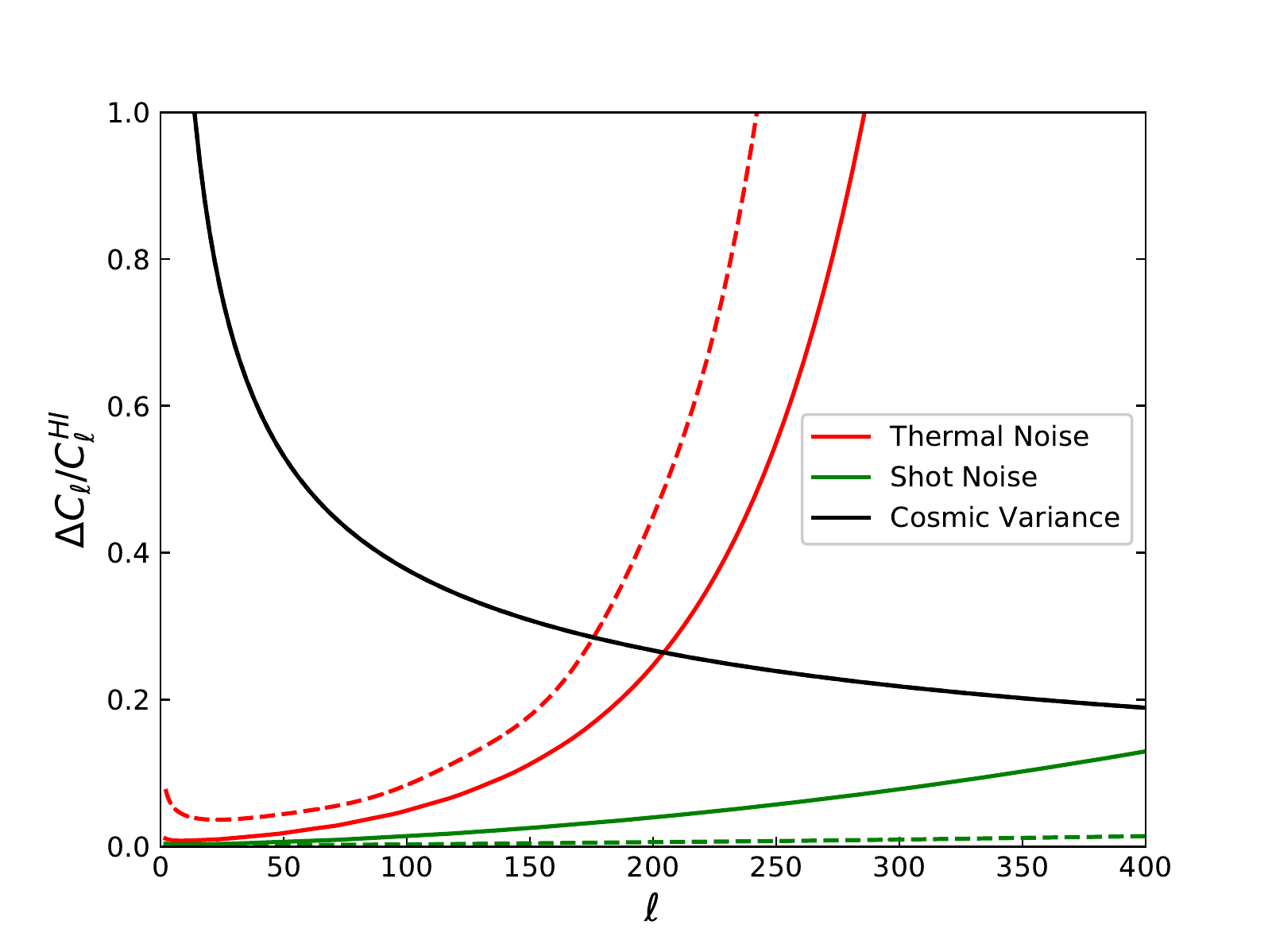}
\caption{\label{Cl_noise} Fractional uncertainties expected for the angular power spectrum for the BINGO experiment ($f_{sky} = 0.07$) for redshifts $z = 0.127$ (solid lines) and $z = 0.449$ (dashed lines) after 1 year of sky integration. The ratio is plotted between different sources of uncertainty (cosmic variance, shot noise, and thermal noise) and the \hi\, angular power spectrum, $\Delta C_\ell / C_\ell^{\textrm{\hi\,}}$, as a function of multipole. For the cosmic variance  the standard formula, $\sqrt{2/(2\ell + 1)f_{\rm{sky}}}$, is used, and for the thermal and shot noise  $\Delta C_\ell$ is equal to their angular power spectrum multiplied by the cosmic variance factor.}
\end{figure}

Due to the discrete nature of the sources emitting an \hi\, signal, the measured auto-spectra have a shot noise contribution in addition to the clustering part described in Sect.~\ref{sec:21cm_Cl}. The shot noise power spectrum can be calculated as \citep{Hall:2012wd}
\begin{equation}
C_{\ell}^{\rm{shot}} = \frac{\bar{T}^2(z)}{\bar{N}(z)} \,,
\end{equation}
where $\bar{N}(z)$ is the angular density of the sources;   assuming a comoving number density $n = \frac{dN}{dV} = 0.03 h^3 $ Mpc$^{-3}$ \citep{Masui:2009cj}, the angular density is  given by
\begin{equation}
\bar{N}(z) = \frac{dN}{d\Omega} = 0.03h^3\textrm{Mpc}^{-3}\frac{c}{H_0}\int{\chi^2(z)\frac{dz}{E(z)}} \,.
\end{equation}
Figure~\ref{Cl_noise} presents these contributions with respect to the \hi\, signal at the first and last redshift bins of BINGO.

\subsection{$1/f$ noise}
In addition to the thermal noise, significant contamination will come from the receiver system caused by gain fluctuations, the detector's temperature changes, quantum fluctuations, and power voltage variations. The power spectrum of this noise is expected to behave as a power law of the frequency. It is generally called  pink noise, or more specifically  $1/f$ noise at low frequencies. It introduces stripes along the scan direction in the observed maps \citep{Bigot-Sazy:2015jaa}, contaminates the \hi\, signal, and can dominate the Gaussian thermal noise. Therefore, $1/f$ noise is an important effect that must be taken into account in order to properly detect the \hi\, signal from single-dish IM radio telescopes.

The power spectral density (PSD) combining both the thermal and $1/f$ noise is the quadratic addition of the two components, given by \citep{Seiffert:2002gh,Bigot-Sazy:2015jaa}
\begin{equation}
    \rm{PSD}(f) = \sigma_T^2\left[1 + \left(\frac{f_{\rm{knee}}}{f}\right)^\alpha\right] \,,
\end{equation}
where $\sigma_T$ is the thermal noise level in Eq.~(\ref{eq:thermal_noise0}), $f_{\rm{knee}}$ is the frequency where the $1/f$ noise has the same amplitude as the thermal noise, and the spectral index $\alpha \approx 1 - 2$ is a parameter. In order to take into account correlations in the frequency direction,  an extension to this equation has been proposed by \cite{Harper:2017gln}
\begin{equation}
   \rm{ PSD}(f, \, \omega) = \sigma_T^2\left[1 + \left(\frac{f_{\rm{knee}}}{f}\right)^\alpha\left(\frac{\omega_0}{\omega}\right)^{\frac{1 - \beta}{\beta}}\right] \,,
\end{equation}
where $\omega = 1/\nu$ is another parameter, ranging from $\omega_0 = (N\Delta\nu)^{-1}$ to $\omega_{N-1} = (\Delta\nu)^{-1}$;  $N$ is the number of frequency channels with width $\Delta\nu$; and $\beta$ is the correlation index that describes the $1/f$ noise correlation in frequency with values in the interval $[0, \, 1]$. The $1/f$ noise will be completely correlated across all frequency channels for $\beta = 0$ and completely uncorrelated if $\beta = 1$.


The effect of $1/f$ noise on 21cm angular power spectra and the final cosmological parameters were analyzed in \citep{Chen:2019jms} for SKA1-MID band 1 and band 2. Here we  extrapolate these results for the case of BINGO. Given the redshift range of BINGO, we expect a degradation from $1/f$ noise that is more similar to what was obtained for SKA1-MID band 2.

In \cite{Chen:2019jms} three cases were considered: with effectively no $1/f$ noise (completely removed by component separation techniques, $\beta = 0$); partially correlated $1/f$ noise ($\beta = 0.5$); and totally uncorrelated $1/f$ noise ($\beta = 1$). The other $1/f$ noise parameters were kept fixed: the slew speed $v_t = 1 \, \textrm{deg s}^{-1}$; the knee frequency $f_{\textrm{knee}} = 1 \, \textrm{Hz}$; and the spectral index $\alpha = 1$. From the area of the $w_0 \times w_a$ joint contour, they found that SKA1-MID band 2 alone was degraded by $\approx 1.5$ with $\beta = 0.5$ and $\approx 2$ with $\beta = 1$. Combining SKA band 2 with \textit{Planck} the degradation was less than a factor of $\approx 1.3$ even at $\beta = 1$.

As discussed in \cite{Chen:2019jms}, higher redshifts and smaller scales are more affected by $1/f$ noise. BINGO will reach lower redshifts than SKA1-MID band 2 and have a better angular resolution, which amplifies the $1/f$ noise as in our thermal noise in Sect.~\ref{sec:tehermal_noise}. Therefore, we expect that the $1/f$ noise will have a degradation factor at most of the same order as found for SKA1-MID band 2 in that paper. Moreover, we aim to obtain with BINGO a knee frequency of $\sim 1\textrm{mHz}$ \citep{2020_instrument}, which would greatly improve our ability to extract the \hi\, signal. In this paper we do not consider further the effect of $1/f$ noise.

\subsection{Foreground residuals}\label{sec:foregorund1}
The success of a 21cm IM experiment will require the effective removal of galactic and extra-galactic foregrounds that can be up to $\sim 10^4$ times stronger than the \hi\, signal.  This requires refined component separation methods able to properly reconstruct the \hi\, signal immersed in the foreground contamination \citep{Olivari:2015tka}. The foreground cleaning process we plan to apply to the BINGO data are presented in the companion papers (\citealt{2020_sky_simulation} and \citealt{2020_component_separation}). In this paper, we assume the foreground cleaning process has already been performed as part of the BINGO pipeline. However, the foreground cleaning procedure leaves some residuals which are also sources of uncertainties. The component separation method GNILC \citep{Remazeilles:2010hq,Remazeilles:2011ze,Olivari:2015tka} projects the observed data into a subspace dominated by \hi\, plus noise and performs an ILC analysis restricted to that space. Therefore, by construction, the foreground residuals should be subdominant. On the other hand, these residuals can introduce a bias in the determination of the 21cm power spectra and, hence, affect our final cosmological constraints.

We can add this bias to our Fisher matrix formalism following the procedure in \cite{Amara:2007as}. For small residual systematics, the bias in the parameter estimation is
\begin{equation}
\label{eq:fisher_bias}
b[\theta_i] = \langle \theta_i \rangle - \langle \theta_i^{\rm{true}} \rangle = \sum_j(F^{-1})_{ij} B_j \,,
\end{equation}
where $\theta_i^{\rm{true}}$ is the true value of the parameters and the bias vector $B_j$ is given by
\begin{equation}
\label{eq:Bj}
B_j = \mathrm{Tr}{\left[{\bf C}^{-1}C_\ell^{\rm{sys}}{\bf C}^{-1}\frac{\partial {\bf C}}{\partial\theta_j}\right]} \,,
\end{equation}
which is similar to the Fisher matrix. In this case  the total error covariance matrix is
\begin{align}
\mathrm{Cov}[\theta_i, \, \theta_j] &= \langle\left(\theta_i - \theta_i^{\rm{true}}\right)\left(\theta_j - \theta_j^{\rm{true}}\right)\rangle \nonumber\\ &= (F^{-1})_{ij} + b[\theta_i]b[\theta_j] \,,
\end{align}
including both statistical and systematical errors. Therefore, the presence of foreground residuals should not only increase the parameter uncertainties, but also shift the center of the error ellipses away from the fiducial model.



\section{Results}
\label{sec:results}
In this section we discuss the expected constraints from the BINGO survey. We adopt the Fisher matrix formalism described in Sect.~\ref{sec:fisher} with the 21cm angular power spectra presented in Sect.~\ref{sec:21cm_Cl}. Unless stated otherwise, we consider an optimal scenario where $1/f$ noise and foreground contamination were already removed using a component separation method (see \citealt{2020_sky_simulation} and \citealt{2020_component_separation}). Therefore, in most of the analysis below we consider the cosmic variance, thermal noise and shot noise only.

First, we  discuss the constraints in the basic $\Lambda$CDM and in a simple extension, the wCDM model. Then, under the scope of the CPL parameterization, we test the effect of several experimental setups on the final cosmological parameters. We consider the effect of varying the number of feed horns, the total observational time, the number of redshift bins, considering or not cross-correlations between redshift bins, and the effect of RSD. In 
Sect.~\ref{sec:foregrounds} we add foreground residuals in our analysis. Finally, in Sect.~\ref{sec:ska} we compare the expected results with SKA band 1 and SKA band 2. Our fiducial experimental setup is given in Table~\ref{tab:bingo_param} and a more detailed description of the instrument can be found in the companion paper II \citep{2020_instrument}.


BINGO will shed light on the \hi\, distribution and evolution at low redshift. In Sect.~\ref{sec:HI_density_bias} we study the expected constraints on the \hi\, density and bias. We also analyze how BINGO can help constraining the total neutrino mass in Sect.~\ref{sec:massive_neutrinos} and alternative cosmologies in Sect.~\ref{sec:alt_cosmo}.

\subsection{The $\Lambda$CDM model}

The $\Lambda$CDM model has been well constrained by the latest CMB measurements made by the \textit{Planck} satellite, with precision of percent to the  sub-percent level on the cosmological parameters \citep{Aghanim:2018eyx}. In this section we  investigate how BINGO can help constrain these parameters.

We  performed the Fisher matrix analysis described in Sect.~\ref{sec:fisher} for BINGO and for  BINGO and  \textit{Planck} combined (BINGO + \textit{Planck}). The results are presented in Table~\ref{tab:LCDM}. In the second column of Table~\ref{tab:LCDM} we note that BINGO alone cannot put competitive constraints on the cosmological parameters from Table~\ref{tab:fiducial_param}. However, the combination of the two surveys can improve the confidence in all cosmological parameters (Column 4 of Table~\ref{tab:LCDM}). The most significant improvements are in the DM density parameter, $\Omega_ch^2$, and the Hubble parameter, $h$, by more than $25\%$ in both of them. This is on the same order as what has been currently obtained by adding CMB lensing and BAO to the final \textit{Planck} temperature and polarization results  \citep[cf. Table 2 in][]{Aghanim:2018eyx}. We can also see that the primordial density parameters, $A_s$ and $n_s$, are mostly constrained by \textit{Planck} itself with an enhancement of $3.8\%$ and $8.7\%$, respectively. The uncertainty on the baryon fraction, parameterized by $\Omega_bh^2$, decreases by $11\%$.
\begin{table*}
\footnotesize
\caption{\label{tab:LCDM} $\Lambda$CDM model. Expected $1\sigma$ constraints on the $\Lambda$CDM cosmological parameters from BINGO, \textit{Planck}, and BINGO + \textit{Planck}. The last column shows the improvement in combining BINGO + \textit{Planck} with respect to the case with \textit{Planck} only, except for $b_{\textrm{\hi\,}}$, which is compared with BINGO only.}
\begin{center}

\sisetup{
        round-mode=figures,
        round-precision=2,
        scientific-notation = fixed,
        fixed-exponent = 0
}

\begin{tabular}{|c|S|S|S|c|}
\cline{2-5}
\multicolumn{1}{c}{}  &
\multicolumn{1}{|c|}{BINGO} & 
\multicolumn{1}{|c|}{\textit{Planck}} & 
\multicolumn{2}{|c|}{BINGO + \textit{Planck} } \\
\hline
 
\multicolumn{1}{|c|}{Parameter} &
\multicolumn{1}{|c|}{$\pm 1\sigma$ ($100\% \times\sigma/\theta_i^{\rm{fid}}$)} &
\multicolumn{1}{|c|}{$\pm 1\sigma$ ($100\% \times\sigma/\theta_i^{\rm{fid}}$)} &
\multicolumn{1}{|c|}{$\pm 1\sigma$ ($100\%\times\sigma/\theta_i^{\rm{fid}}$)} &
\multicolumn{1}{|c|}{$100\%\times|\sigma_{\textrm{total}} - \sigma_{\textrm{ref}}|/\sigma_{\textrm{ref}}$} \\
 
 \hline
$\Omega_b h^2$ & 0.01411938 \textrm{  (63\%)} & 0.00014962 \textrm{  (0.7\%)} & 0.00013266 \textrm{  (0.6\%)}  & 11\% \\
$\Omega_c h^2$ & 0.04522356 \textrm{  (38\%)} & 0.00136289 \textrm{  (1.1\%)} & 0.00101731 \textrm{  (0.8\%)} & 25\% \\
$h$ & 0.12940198 \textrm{  (19\%)} & 0.00605864 \textrm{  (0.9\%)} & 0.00448605 \textrm{  (0.7\%)} & 26\% \\
${\rm{ln}}(10^{10}A_s)$ & 0.67966409 \textrm{  (22\%)} & 0.01570926 \textrm{  (0.5\%)} & 0.0151182 \textrm{  (0.5\%)} & 3.8\% \\
$n_s$ & 0.08879796 \textrm{  (9.2\%)} &  0.0043047 \textrm{  (0.4\%)} & 0.00392966 \textrm{  (0.4\%)} & 8.7\% \\
$b_{\textrm{\hi\,}}$ & 0.04091672 \textrm{  (4.1\%)} &   & 0.0105321 \textrm{  (1.1\%)} & 74\%  \\
 \hline
\end{tabular}
\end{center}

\end{table*}

On the other hand, the most important contribution from 21cm experiments will not be  reducing these error bars, but providing additional information from a different tracer of the matter distribution. Although the $\Lambda$CDM model has been in good agreement with data, recent observations have pointed out severe tensions between the CMB and low redshift data under the $\Lambda$CDM scenario.  \cite{Riess:2019cxk} obtained a $4.4\sigma$ deviation of the value for the Hubble constant measured by \textit{Planck} and the one using Cepheids in the Large Magellanic Cloud. Measurements from weak gravitational lensing have also pointed out a disagreement of $2.3\sigma$ for the value of $S_8=\sigma_8\sqrt{\Omega_m/0.3}$ obtained with \textit{Planck} data \citep{Hildebrandt:2018yau}. Therefore, a new and independent tracer of the matter distribution, with different systematics, will provide valuable information and improve our understanding of the evolution of the  Universe. Our projected constraints show that BINGO alone will not have enough power to solve these tensions; however, it can be combined with other constraints and will also open the way to more precise 21cm experiments in the future.

\subsection{wCDM model}
Because of the extra freedom in the parameter space, the previous constraint from \textit{Planck} data in the Hubble constant is loosened. The $1\%$ precision measurement for the Hubble constant at $1\sigma$ in the $\Lambda$CDM model goes to $\sim 13\%$ in the wCDM model. The ability to constrain the DE EoS using CMB information only is also rather weak, providing $25\%$ uncertainty at a $1\sigma$ confidence level (CL). In this case additional data at low redshifts where DE dominates is crucial.
\begin{table*}
\footnotesize
\caption{\label{tab:wCDM} wCDM model. Expected $1\sigma$ constraints on the wCDM cosmological parameters from BINGO, \textit{Planck}, and BINGO + \textit{Planck}. The last column shows the improvement in combining BINGO + \textit{Planck} with respect to the case with \textit{Planck} only, except for $b_{\textrm{\hi\,}}$, which is compared with BINGO only.}
\begin{center}

\sisetup{
        round-mode=figures,
        round-precision=2,
        scientific-notation = fixed,
        fixed-exponent = 0
}

\begin{tabular}{|c|S|S|S|c|}
\cline{2-5}
\multicolumn{1}{c}{}  &
\multicolumn{1}{|c|}{BINGO} & 
\multicolumn{1}{|c|}{\textit{Planck}} & 
\multicolumn{2}{|c|}{BINGO + \textit{Planck} } \\
\hline
 
\multicolumn{1}{|c|}{Parameter} &
\multicolumn{1}{|c|}{$\pm 1\sigma$ ($100\%\times\sigma/\theta_i^{\rm{fid}}$)} &
\multicolumn{1}{|c|}{$\pm 1\sigma$ ($100\%\times\sigma/\theta_i^{\rm{fid}}$)} &
\multicolumn{1}{|c|}{$\pm 1\sigma$ ($100\%\times\sigma/\theta_i^{\rm{fid}}$)} &
\multicolumn{1}{|c|}{$100\%\times|\sigma_{\textrm{total}} - \sigma_{\textrm{ref}}|/\sigma_{\textrm{ref}}$} \\
 
 \hline
$\Omega_b h^2$ & 0.01418054 \textrm{  (63\%)} & 1.50991477e-04 \textrm{  (0.7\%)} & 0.00013869 \textrm{  (0.6\%)} & 8.1\% \\
$\Omega_c h^2$ & 0.04576505 \textrm{  (38\%)} & 1.40238664e-03 \textrm{  (1.2\%)} & 0.0011141 \textrm{  (0.9\%)} & 21\% \\
$h$ & 0.13193422 \textrm{  (20\%)} & 8.86369432e-02 \textrm{  (13\%)} & 0.00731426 \textrm{  (1.1\%)} & 92\% \\
${\rm{ln}}(10^{10}A_s)$ & 0.73213243 \textrm{  (24\%)} & 1.60059411e-02 \textrm{  (0.5\%)} & 0.01589729 \textrm{  (0.5\%)} & 0.7\% \\
$n_s$ & 0.09020935 \textrm{  (9.3\%)} & 4.38944062e-03 \textrm{  (0.5\%)} & 0.00401808 \textrm{  (0.4\%)} & 8.5\% \\
$w_0$ & 0.1700468 \textrm{  (17\%)} & 2.55371116e-01 \textrm{  (25\%)} & 0.03286446 \textrm{  (3.3\%)} & 87\% \\
$b_{\textrm{\hi\,}}$ & 0.05229512 \textrm{  (5.2\%)} &   & 0.02261605 \textrm{  (2.3\%)} & 57\%  \\
 \hline
\end{tabular}
\end{center}

\end{table*}

\begin{figure}[]
\centering
\includegraphics[scale=0.6]{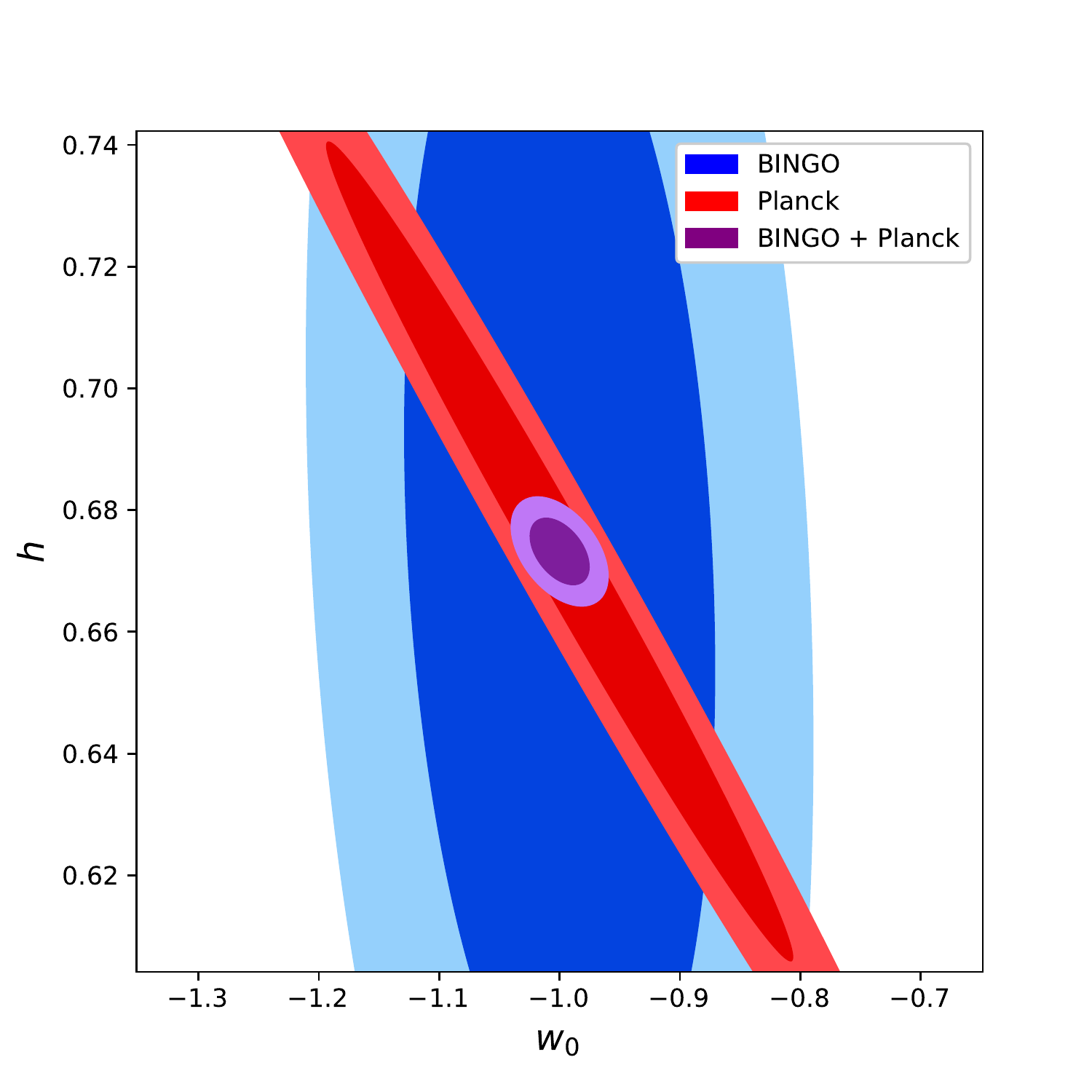}
\caption{\label{fig:wCDM}Marginalized constraints ($68\%$ and $95\%$ CL)    on the DE EoS and Hubble parameters for the wCDM model using BINGO, \textit{Planck}, and BINGO + \textit{Planck}.}
\end{figure}

The 21cm IM power spectra measured by BINGO will be able to improve these constraints. In Table~\ref{tab:wCDM} we show the expected constraints in the wCDM model by BINGO and in combination with \textit{Planck}. We can see that BINGO alone will be able to put similar constraints to those from  \textit{Planck} on the Hubble constant and a $17\%$ precision measurement at $1\sigma$ CL on the DE EoS. Combining them  can greatly improve these results,   reaching a remarkable precision of $1.1\%$ for $H_0$ and $3.3\%$ for the EoS, which are consistent with previous results obtained by \cite{Olivari:2017bfv}. The reason for this improvement can be observed in Fig.~\ref{fig:wCDM}. BINGO can help break the degeneracy between $H_0$ and $w$ in the \textit{Planck} data, dramatically improving the constraints.

As a comparison, recent measurements from the three years of the Dark Energy Survey (DES) \citep{DES:2021wwk} using three two-point correlation functions (from cosmic shear, galaxy clustering, and cross-correlation of source galaxy shear with lens galaxy positions) over $5000 \, \textrm{deg}^2$ obtained a constraint on the DE EoS of $w = -0.98^{+0.32}_{-0.20}$. Combining their measurement with \textit{Planck} (no lensing), their marginalized constraint is given by $w = -1.090^{+0.128}_{-0.113}$. Therefore, although we are still performing an optimistic analysis, we can see that BINGO has the potential to put  constraints that is competitive with the current LSS surveys.

\subsection{CPL parameterization}

\begin{figure}[h!]
\centering
\includegraphics[scale=0.6]{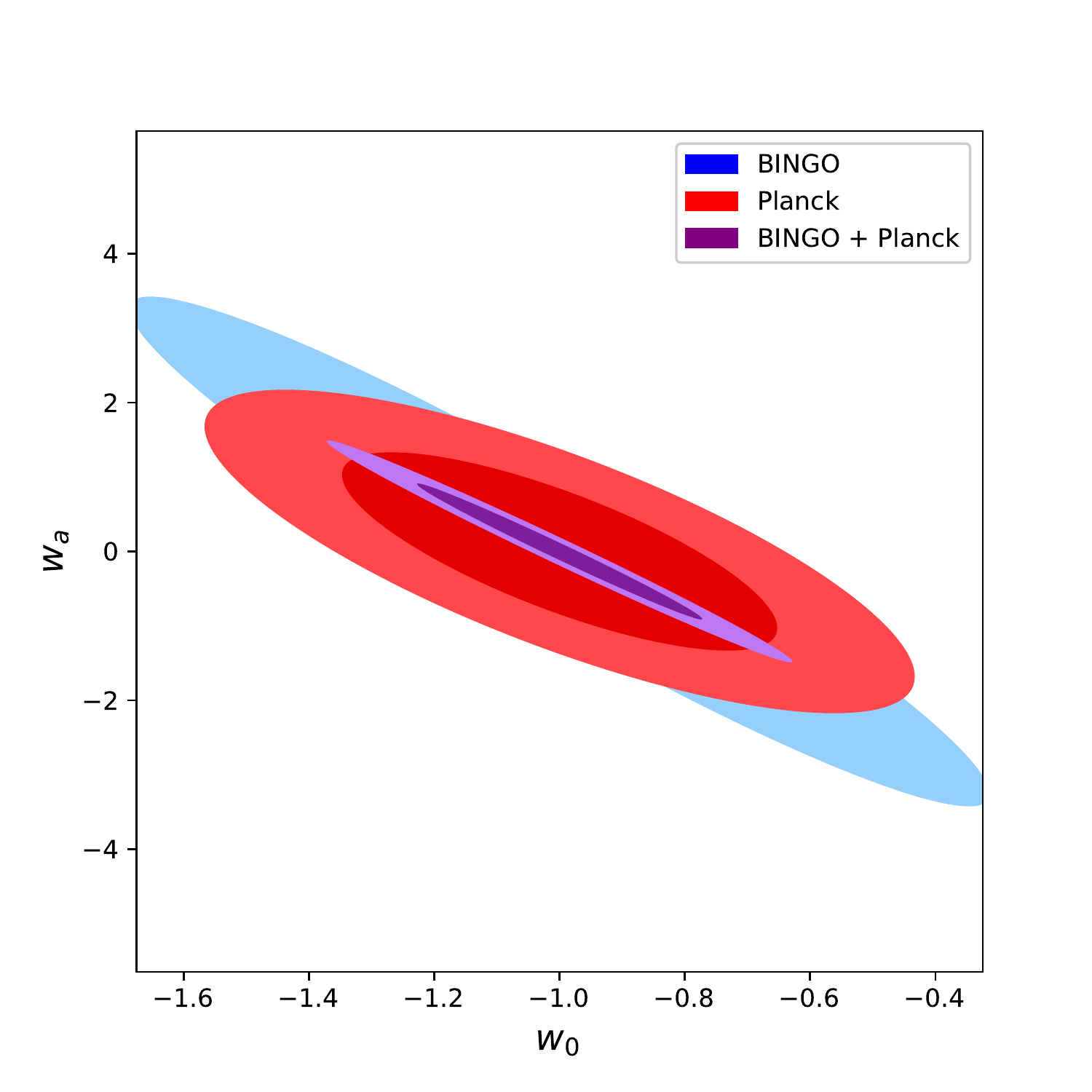}
\caption{\label{fig:plot_w_wa} Marginalized constraints ($68\%$ and $95\%$ CL)   on the DE EoS parameters for the CPL parameterization using BINGO, \textit{Planck}, and BINGO + \textit{Planck}.}
\end{figure}

Considering the BINGO fiducial setup, we show the forecast constraints on the CPL parameterization in Table~\ref{tab:CPL}. The additional parameter  increased the uncertainties on the cosmological variables, as expected. The combination of 21cm IM from BINGO with CMB data from \textit{Planck} can put a $2.9\%$ constraint on the Hubble constant, $30\%$ on the DE EoS parameter $w_0$, and a $\sigma_{w_a}$ = 1.2 at $68\%$ CL. Although these constraints are still large, BINGO has improved the results from \textit{Planck} alone by $78\%$ for $H_0$, $34\%$ for $w_0$, and $31\%$ for $w_a$. Figure~\ref{fig:plot_w_wa} shows the $68\%$ and $95\%$ confidence contours in the $w_0 \times w_a$ parameter space.
\begin{table*}
\footnotesize
\caption{\label{tab:CPL} CPL model. Expected $1\sigma$ constraints on the CPL cosmological parameters from BINGO, \textit{Planck}, and BINGO + \textit{Planck}. The last column shows the improvement in combining BINGO + \textit{Planck} with respect to the case with \textit{Planck} only, except for $b_{\textrm{\hi\,}}$, which is compared with BINGO only.}
\begin{center}

\sisetup{
        round-mode=figures,
        round-precision=2,
        scientific-notation = fixed,
        fixed-exponent = 0
}

\begin{tabular}{|c|S|S|S|c|}
\cline{2-5}
\multicolumn{1}{c}{}  &
\multicolumn{1}{|c|}{BINGO} & 
\multicolumn{1}{|c|}{\textit{Planck}} & 
\multicolumn{2}{|c|}{ BINGO + \textit{Planck}} \\
\hline
 
\multicolumn{1}{|c|}{Parameter} &
\multicolumn{1}{|c|}{$\pm 1\sigma$ ($100\% \times\sigma/\theta_i^{\rm{fid}}$)} &
\multicolumn{1}{|c|}{$\pm 1\sigma$ ($100\%\times\sigma/\theta_i^{\rm{\rm{fid}}}$)} &
\multicolumn{1}{|c|}{$\pm 1\sigma$ ($100\%\times\sigma/\theta_i^{\rm{\rm{fid}}}$)} &
\multicolumn{1}{|c|}{$100\%\times|\sigma_{\textrm{total}} - \sigma_{\textrm{ref}}|/\sigma_{\textrm{ref}}$} \\
 
 \hline
$\Omega_b h^2$ & 0.01527505 \textrm{  (68\%)} & 1.55548076e-04 \textrm{  (0.7\%)} & 1.44788222e-04 \textrm{  (0.6\%)} & 6.9\% \\
$\Omega_c h^2$ & 0.05178439 \textrm{  (43\%)} & 1.34573801e-03 \textrm{  (1.1\%)} & 1.09590859e-03 \textrm{  (0.9\%)} & 19\% \\
$h$ & 0.13509788 \textrm{  (20\%)} & 8.81208091e-02 \textrm{  (13\%)} & 1.93071665e-02 \textrm{  (2.9\%)} & 78\% \\
${\rm{ln}}(10^{10}A_s)$ & 0.92028096 \textrm{  (30\%)} & 1.57720075e-02 \textrm{  (0.5\%)} & 1.56795671e-02 \textrm{  (0.5\%)} & 0.6\% \\
$n_s$ & 0.10813622 \textrm{  (11\%)} & 4.43100451e-03 \textrm{  (0.5\%)} & 4.07617674e-03 \textrm{  (0.4\%)} & 8.0\% \\
$w_0$ & 0.55395363 \textrm{  (55\%)} & 4.57265951e-01 \textrm{  (46\%)} & 2.99878642e-01 \textrm{  (30\%)} & 34\% \\
$w_a$ & 2.76061207  & 1.75186826e+00 & 1.20179533e+00 & 31\% \\
$b_{\textrm{\hi\,}}$ & 0.08050553 \textrm{  (8.1\%)} &   & 2.26292876e-02 \textrm{  (2.3\%)} & 72\%  \\
 \hline
\end{tabular}
\end{center}

\end{table*}

\cite{Yohana2019} have  made a forecast for BINGO using the angular power spectra under the CPL parameterization. Their analysis considers the same cosmological parameters used here plus the effective number of relativistic neutrinos, $N_{\textrm{eff}}$, and the sum of neutrino masses, $\sum m_\nu$. The \hi\, bias was kept fixed. Their projected constraints for the DE EoS parameters were significantly weaker than ours, while for the Hubble constant it was more than two times stronger. Although the degradation in the DE EoS parameters can be understood in terms of the extra degrees of freedom, the difference in the Hubble constant may be related to differences in the analysis and BINGO setup.

\subsubsection{Effect of total observational time}
\begin{figure}[h!]
\centering
\includegraphics[scale=0.6]{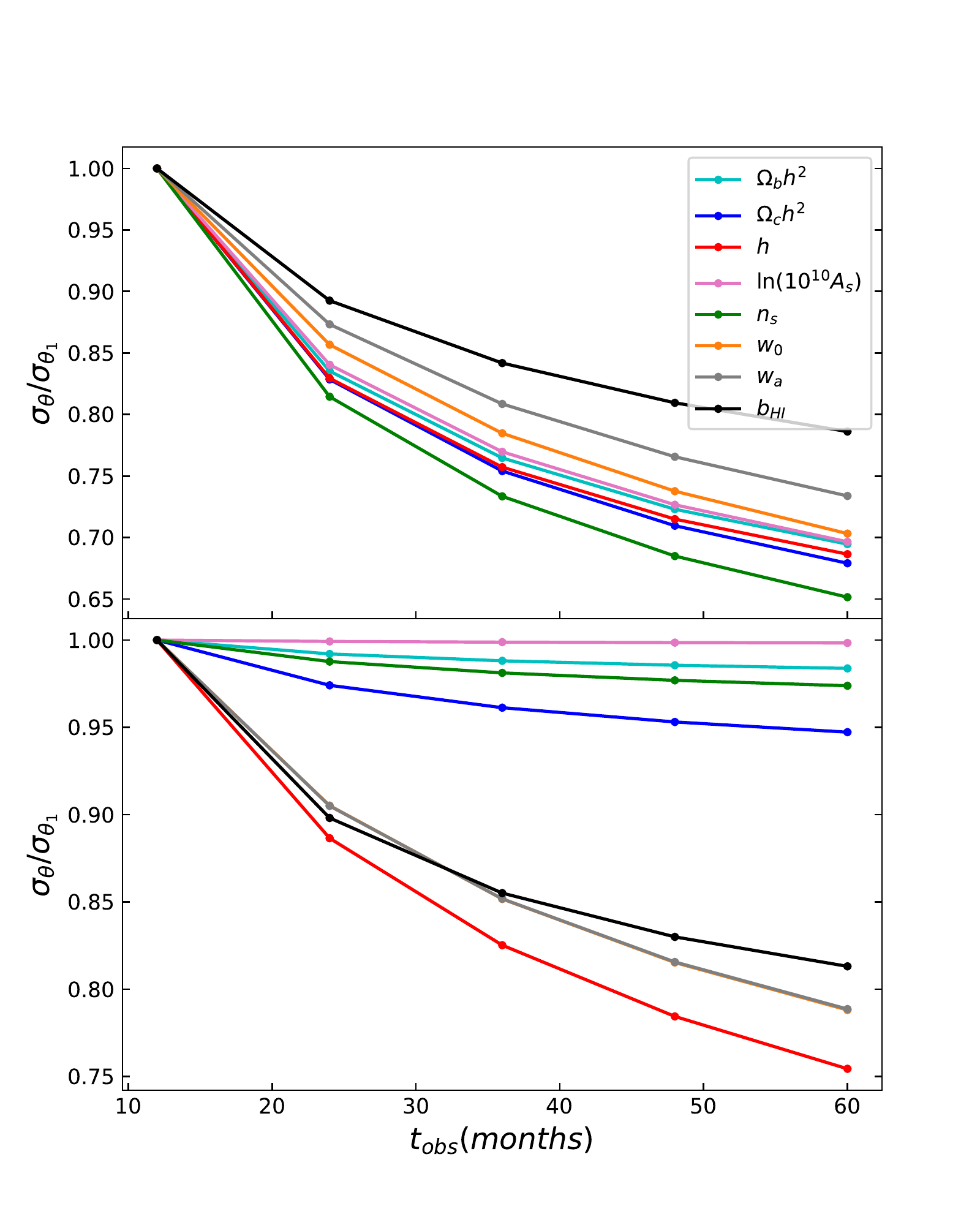}
\caption{\label{plot_tobs}Constraints on the cosmological and \hi\, parameters as a function of the total observational time relative to those with $t_{obs} = 1$ year for BINGO (top) and BINGO + \textit{Planck} (bottom). We vary $t_{obs}$ in a range of $[1, \, 2, \, 3,\, 4,\, 5]$ years. {In the bottom plot $w_a$ is on top of $w_0$, which cannot be seen.} }
\end{figure}

Using the CPL parameterization as our fiducial cosmology, we analyze how different experimental setups can impact our final measurements of cosmological parameters. We start by considering the impact of BINGO's total observational time. Figure~\ref{plot_tobs} presents the results for 1, 2, 3, 4, and 5 years of the BINGO survey. We show the results for BINGO only   and in combination with \textit{Planck}. Considering BINGO alone, a five-year experiment can improve the constraints in a range from $21\%$ to $35\%$. We can observe that the inclination of these curves decreases, going to a plateau, but it has not yet been achieved at five years.

In order to better evaluate the improvement in our parameter space, we calculate the figure of merit (FoM), defined as the volume of the error ellipsoid $\textrm{FoM} \equiv V \propto \textrm{det} (F)^{-1/2}$. Table~\ref{tab:fom} presents these values for BINGO and BINGO + {\it Planck} as a function of the total observational time. We can observe that the ellipsoid volume decreases by 11 times from 1 year to 5 years with BINGO only, and significantly flattens after year 3.
\begin{table}
\footnotesize
\caption{\label{tab:fom} Figure of merit as a function of the total observational time for BINGO and BINGO + \textit{Planck} under the CPL parameterization.}
\begin{center}


\begin{tabular}{|c|c|c|}
\cline{2-3}
\multicolumn{1}{c}{}  &
\multicolumn{2}{|c|}{$\textrm{FoM} \equiv \textrm{det} (F)^{-1/2} $} \\
\hline
 
\multicolumn{1}{|c|}{$t_{\rm{obs}}$} &
\multicolumn{1}{|c|}{BINGO} &
\multicolumn{1}{|c|}{BINGO + \textit{Planck}} \\
 
 \hline
1 year &  $2.6 \times 10^{-12}$  &  $4.5 \times 10^{-18}$  \\
2 years &  $8.2 \times 10^{-13}$  &  $2.8 \times 10^{-18}$  \\
3 years &  $4.6 \times 10^{-13}$  &  $2.2 \times 10^{-18}$  \\
4 years &  $3.1 \times 10^{-13}$  &  $1.9 \times 10^{-18}$  \\
5 years &  $2.4 \times 10^{-13}$  &  $1.7 \times 10^{-18}$  \\
 \hline
\end{tabular}
\end{center}

\end{table}

On the other hand, the combination with \textit{Planck} data shows that some parameters are not strongly dependent on the BINGO setup since they are mostly constrained by \textit{Planck}. BINGO will mainly affect the measurement of the Hubble constant, DE EoS parameters, and the \hi\, bias, as has already been observed \citep{Olivari:2017bfv}. A five-year experiment can improve the bias measurement by $19\%$, the EoS parameters ($w_0$ and $w_a$) by about $21\%$ each, and the Hubble constant by $25\%$ compared to the standard case. Combining the two surveys, the error ellipsoid decreases by 2.7 times from 1 to 5 years of IM survey.

\subsubsection{Effect of varying the number of horns}
\begin{figure}[h!]
\centering
\includegraphics[scale=0.6]{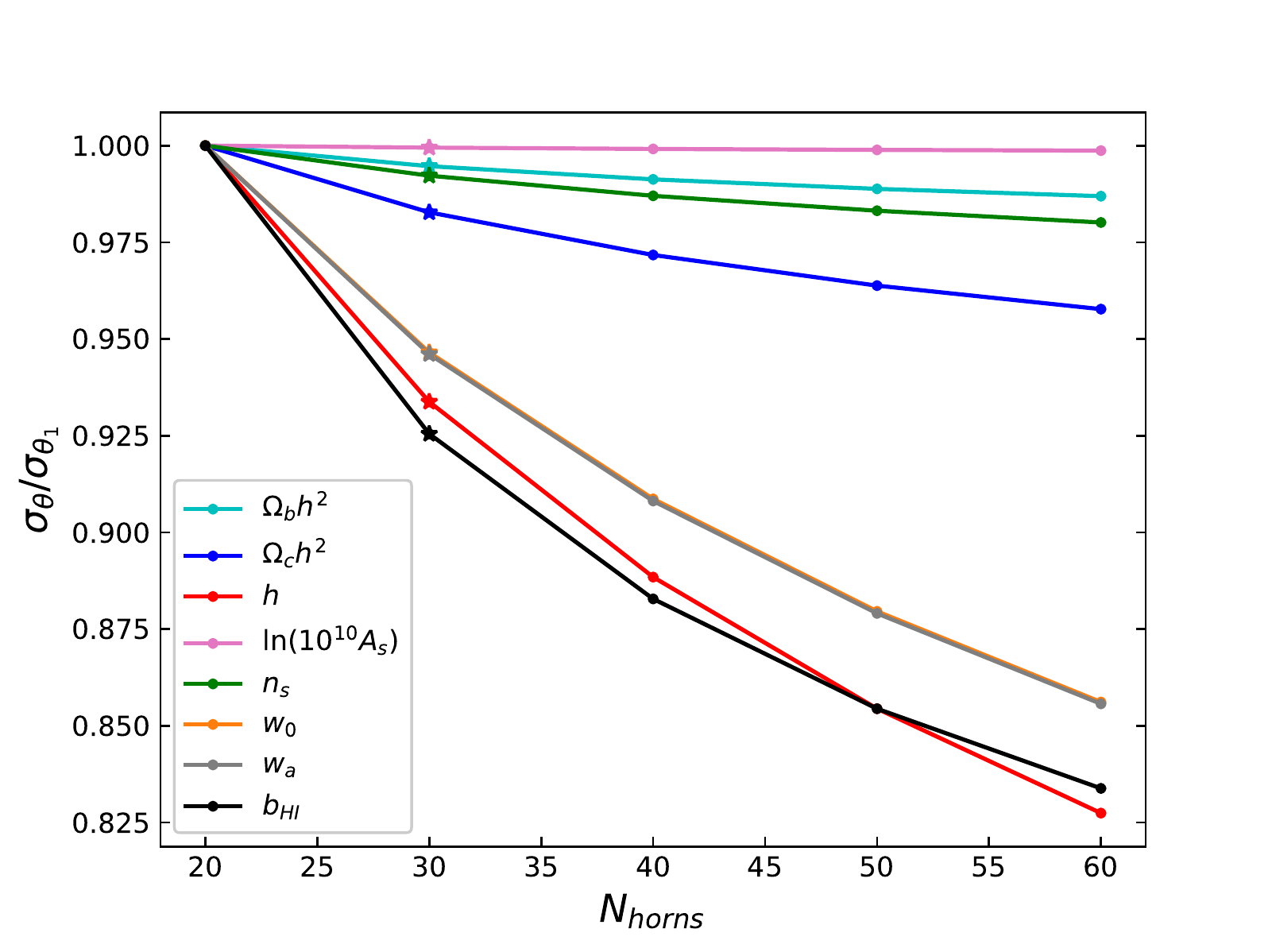}
\caption{\label{plot_horns}Constraints on the cosmological and \hi\, parameters as a function of the number of feed horns for BINGO + \textit{Planck}.}
\end{figure}

In this section we consider the effect of the total number of feed horns. As described in Table~\ref{tab:bingo_param}, the BINGO standard setup will consist of 28 feed horns. In Fig.~\ref{plot_horns} we represent the BINGO + \textit{Planck} standard scenario by a star. Then, keeping all parameters fixed and only varying the number of horns as $N_{\rm{horns}} = 20, 30, 40, 50, 60$, we observe its effect on the cosmological constraints. We observe that some parameters are mainly constrained by \textit{Planck} and, therefore, will not be very sensitive to BINGO's number of horns. The most affected parameters are the DE EoS parameters, the Hubble constant, and the \hi\, bias. They improve by about $15\%$ with respect to $N_{\rm{horns}} = 20$, more specifically $\delta{w_0}$ and $\delta{w_a} = 14\%$, $\delta{b_{\textrm{\hi\,}}} = 16\%$, and $\delta{h} = 17\%$.

We observe, however, that these curves are only taking into account how the number of horns affects the thermal noise without changing the total observational area. A simple telescope design with drift scan only may not be able to cover the same fraction of the sky if the number of horns is too small. Several horn arrangements are discussed in the BINGO companion paper \citep{2020_optical_design}:  a rectangular configuration ($n_f = 33$), a double-rectangular configuration ($n_f = 28$), and a hexagonal configuration ($n_f = 49$). In all cases the total surveyed area was kept constant. In this case, as can be observed in Eq.~(\ref{eq:thermal_noise}), varying the total observational time or the number of horns produces the same result if they change accordingly.

\subsubsection{Effect of the number of redshift bins}
\label{sec:bins}
Our analysis of the volume surveyed by BINGO was   done using the 21cm angular power spectra in redshift bins. The angular power spectrum projects all contributions inside a thin shell. Therefore, we can make a tomographic analysis of the Universe volume. If we sliced that volume into thinner shells, we could obtain a more detailed observation; however, as can be observed in Eq.~(\ref{eq:thermal_noise}), thinner shells also mean larger thermal noise. Therefore, we expect an optimal number of slices beyond which no more cosmological information could be extracted from a specific survey. In addition, the number of necessary calculations to take all auto- and cross-correlated $C_\ell$s into account increases as we increase the number of redshift bins, hence it is desirable to keep this number as small as possible for computational purposes.

\begin{figure}[h!]
\centering
\includegraphics[scale=0.6]{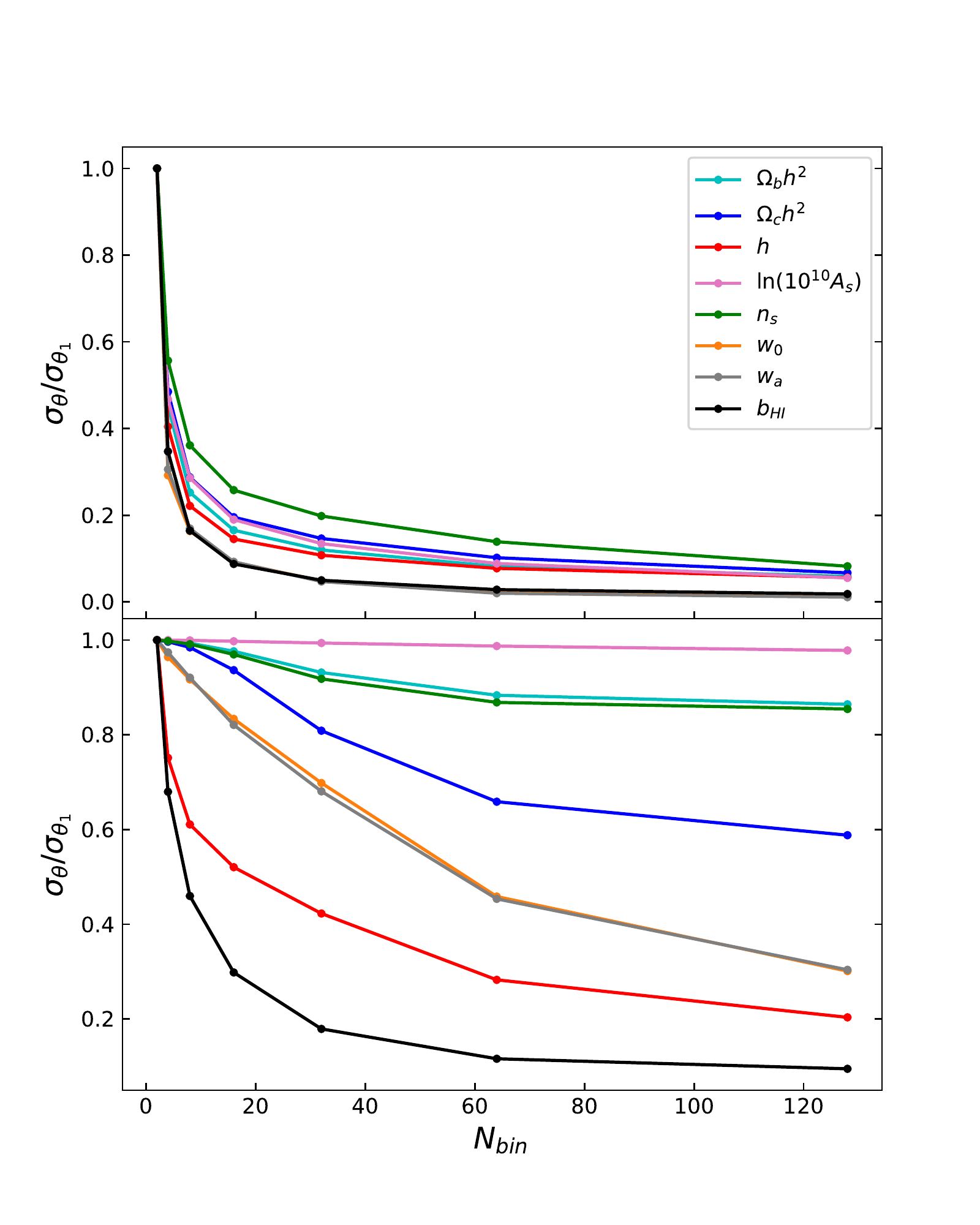}
\caption{\label{plot_bins} Projected constraints on the cosmological and \hi\, parameters as a function of the number of bins for BINGO (top) and BINGO + \textit{Planck} (bottom). They show that $w_0$, $w_a$, and $h$ can have their constraints further improved for $N_{\rm{bin}}$ > 30.}
\end{figure}

We study this  behavior with the BINGO telescope considering $N_{\rm{bin}} = 2, 4, 8, 16, 32, 64, 128$, which implies in equally spaced frequency bins with bandwidths equal to $\Delta\nu = 140, 70, 35, 17.5, 8.75, 4.375, 2.187$ MHz. Figure~\ref{plot_bins} presents our results for BINGO  and for  BINGO + \textit{Planck}.  We can see that increasing the number of bins can greatly improve several cosmological constraints, especially those related to the late-time cosmic acceleration. On the other hand, there is not much difference from $N_{\rm{bin}} = 64$ to $N_{\rm{bin}} = 128$ and    a plateau has been reached  for several parameter uncertainties.
The projected uncertainties with BINGO for $N_\textrm{bin} = 128$ relative to $N_\textrm{bin} = 2$ have improved by about $92$ times for $w_a$, $74$ times for $w_0$, $55$ times for $b_{\textrm{\hi\,}}$, $18$ times for $\mathrm{ln}(10^{10}A_s)$ and $h$, $17$ times for $\Omega_bh^2$, $15$ times for $\Omega_ch^2$, and $12$ times for $n_s$. On the other hand, if we compare our results for $N_{\textrm{bin}} = 128$ with $N_{\textrm{bin}} = 64$, the maximum improvement is of $1.8$ times for $w_a$. The comparison with $N_{\textrm{bin}} = 32$, which is close to our  fiducial value, shows that $N_{\textrm{bin}} = 128$ improves our constraints by about $4$ times for the DE EoS parameters $w_0$ and $w_a$, $3$ times for $b_{\textrm{\hi\,}}$, and $2$ times for the other parameters. In the case of BINGO + {\it Planck}, the constraints with $N_{\rm{bin}} = 128$ are smaller than those with $N_{\rm{bin}} = 2$ by about $11$ times for $b_{\textrm{\hi\,}}$, $5$ times for $h$, $3$ times for $w_0$ and $w_a$, $1.7$ times for $\Omega_ch^2$, $1.2$ times for $\Omega_bh^2$ and $n_s$, and is negligible for $\mathrm{ln}(10^{10}A_s)$. Comparing $N_{\rm{bin}} = 128$ with $N_{\rm{bin}} = 64$, the largest improvements are $1.5$ times for the DE EoS parameters, followed by the Hubble constant with an uncertainty that is  $1.4$ times smaller.

\subsubsection{Effect of cross-correlations}
\begin{figure}[h!]
\centering
\includegraphics[scale=0.6]{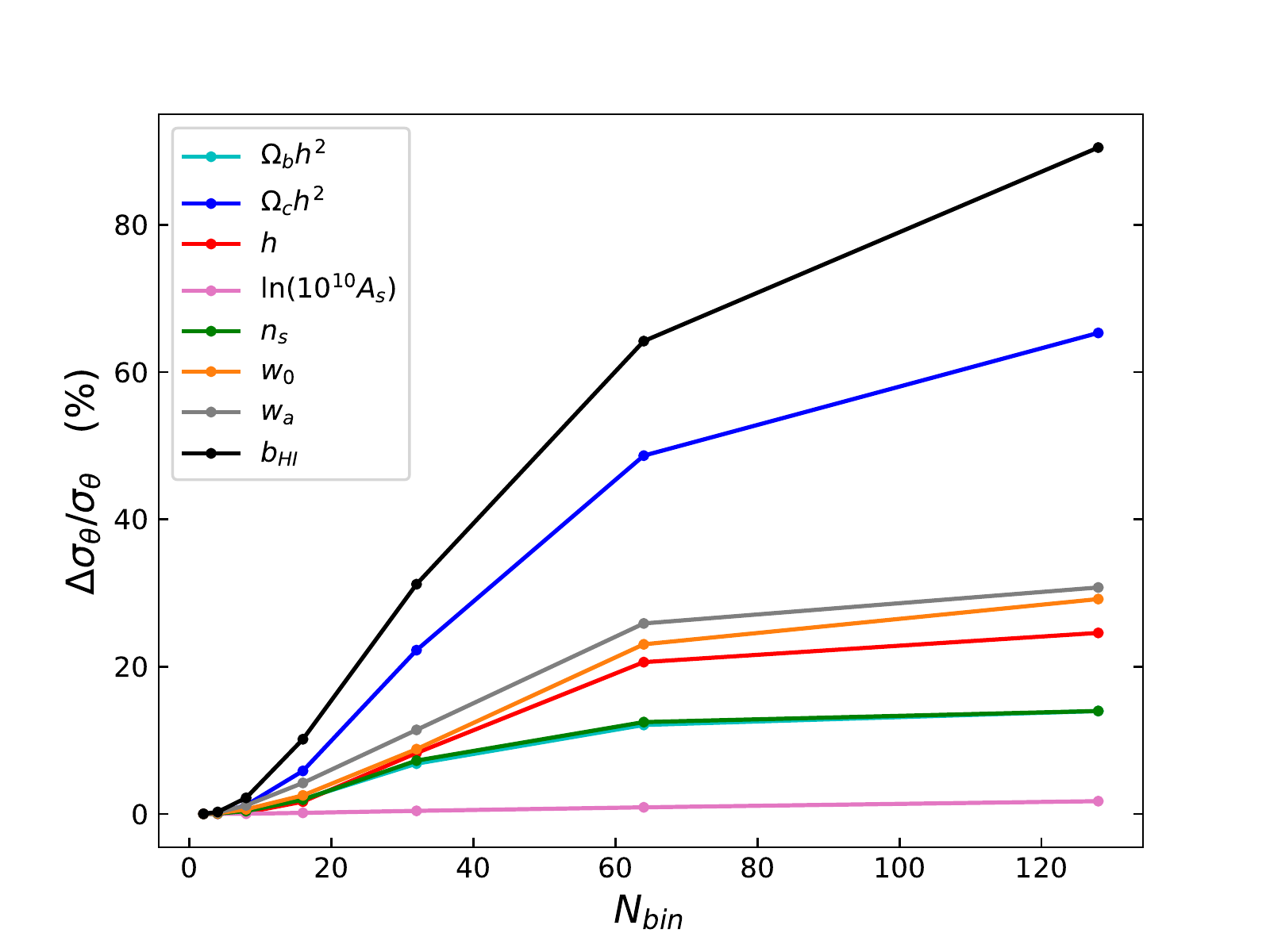}
\caption{\label{plot_no_corr1} Percentage difference between using and not using information from the cross-correlations in the projected cosmological parameter constraints expected for BINGO + \textit{Planck}.}
\end{figure}

{Previous analyses} in the literature considered the Limber approximation to forecast 21cm IM constraints \citep{Olivari:2017bfv}. The Limber approximation does not take into account the cross-correlations between redshift bins, only the auto-correlation spectra. Although this allows a faster and simpler calculation, we miss part of the information contained in the whole spectra. Here we study the effect  including all information using the full power spectra. We demonstrate that effect in Fig.~\ref{plot_no_corr1}, where we calculate the percentage difference between the results with and without cross-correlation ($\Delta \sigma_{\theta}/\sigma_{\theta} =\sigma_{\theta\,\text{without} }/\sigma_{\theta\, \text{with} } - 1 $) as a function of the number of bins. As  can be seen, the importance of cross-correlations increases as we increase the number of bins, but it eventually  reaches a plateau. For $N_{\rm{bin}} = 32$, which is close to the fiducial BINGO setup, we find that the cross-correlations improve the constraints by $\delta{\mathrm{ln}(10^{10}A_s)} = 0.4\%$, $\delta{\Omega_bh^2} = 6.8\%$, $\delta{n_s} = 7.2\%$, $\delta{h} = 8.3\%$, $\delta{w_0} = 8.8\%$, $\delta{w_a} = 11\%$, $\delta{\Omega_ch^2} = 22\%$, and $\delta{b_{\textrm{\hi\,}}} = 31\%$. The primordial spectrum amplitude, $A_s$, is the least affected by cross-correlations as it is basically constrained by \textit{Planck} data. At the maximum number of bins considered, these constraints have improved to $\delta{\mathrm{ln}(10^{10}A_s)} = 1.7\%$, $\delta{\Omega_bh^2}$ and $\delta{n_s} = 14\%$, $\delta{h} = 25\%$, $\delta{w_0} = 29\%$, $\delta{w_a} = 31\%$, $\delta{\Omega_ch^2} = 65\%$, and $\delta{b_{\textrm{\hi\,}}} = 90\%$.

\subsubsection{Effect of RSD}\label{sec:rsd}
\begin{figure}[h!]
\centering
\includegraphics[scale=0.6]{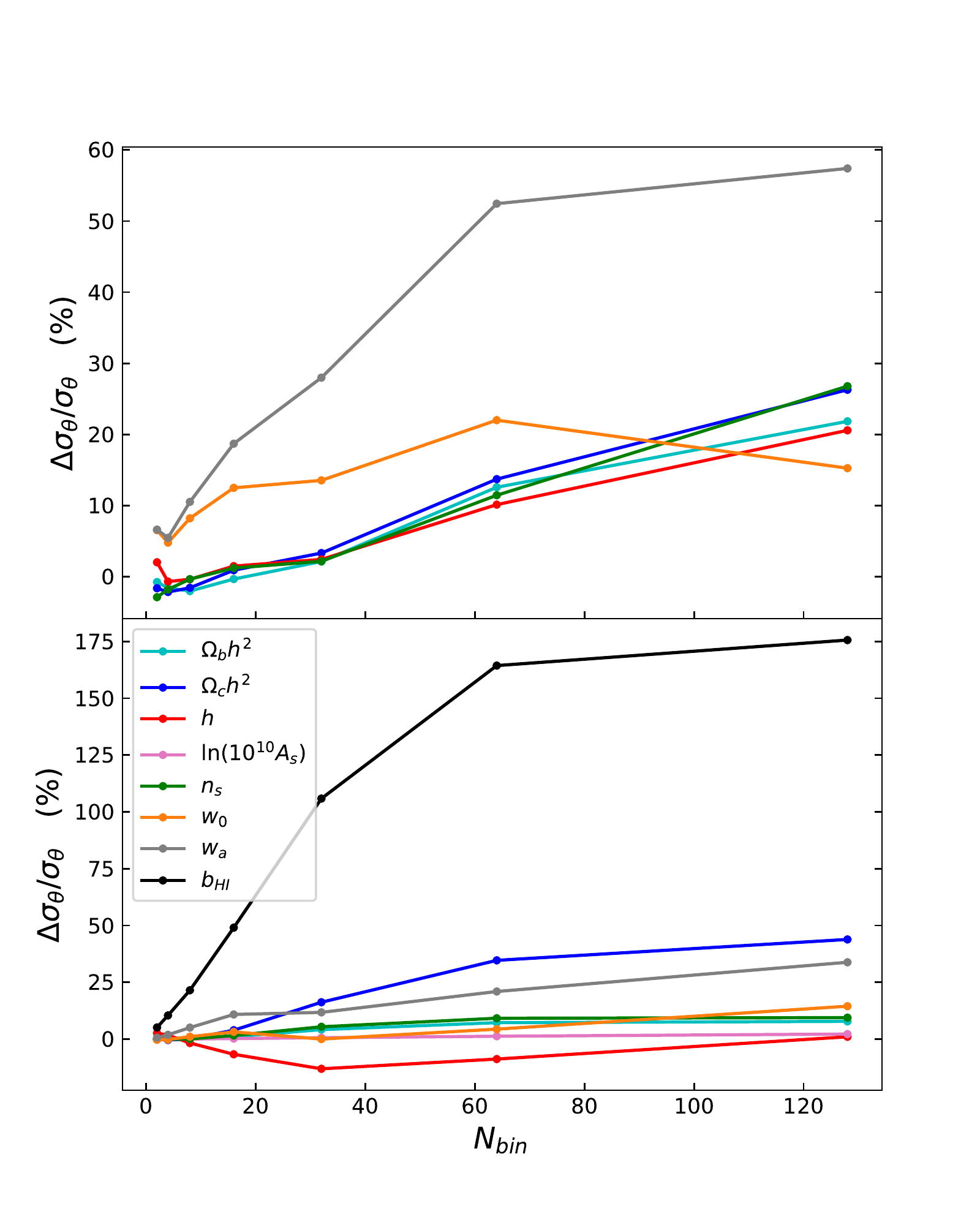}
\caption{\label{plot_rsd} Percentage difference between using and not using information from RSD in the cosmological parameter constraints from BINGO (top) and BINGO + \textit{Planck} (bottom). }
\end{figure}

Another feature that is worth investigating is the effect of RSD. First, we observe that without RSD the primordial scalar amplitude and the \hi\, bias are completely degenerate. RSD is able to break that degeneracy and allows us to constrain these parameters individually with 21cm data alone. We show the percentage difference between the Fisher matrix results with and without RSD ($\Delta \sigma_{\theta}/\sigma_{\theta} =\sigma_{\theta\,\text{without} }/\sigma_{\theta\, \text{with} } - 1 $) in Fig.~\ref{plot_rsd}. The parameters $A_s$ and $b_{\textrm{\hi\,}}$ go from a complete ignorance to constraints on the order of the percent level, and therefore we do not include them in the plot with BINGO alone. Second, we expect  RSD to become more and more important as the frequency (or redshift) smoothing width gets narrower. This happens because the angular spectra sum up all contributions inside the bin, and hence a large bandwidth will cancel out the contributions. This behavior can be seen in Fig.~\ref{plot_rsd}, where there is a tendency for RSD to improve the cosmological constraints as we decrease the bandwidth.

Considering BINGO alone, the  most significantly affected parameters by RSD at $N_{\rm{bin}} = 32$ are the EoS parameters, with $\delta w_0 = 13\%$ and $\delta w_a = 30\%$. The other parameters improve by at most $\sim 3\%$. Increasing the number of bins, we achieve a difference between using and not using RSD in a range from $\sim 15\%$ to $\sim 57\%$ at $N_{\rm{bin}} = 128$. If we combine our \hi\, results with CMB data from \textit{Planck}, the CMB measurements can put constraints on the amplitude of the primordial spectrum, $A_s$, and break the degeneracy with our \hi\, bias even if we do not consider RSD. Figure~\ref{plot_rsd} also includes the results in combination with CMB data. The addition of \textit{Planck} data will put tight constraints on and will affect the correlation between several parameters; therefore, RSD from our 21cm IM spectra will behave differently from the earlier results. We can observe this in the behavior of the Hubble constant parameter, which increases the uncertainty with the inclusion of RSD for several values of redshift bins. At $N_{\rm{bin}} = 128$ the improvements from RSD are given by $\delta{h} = 1\%$, $\delta{\mathrm{ln}(10^{10}A_s)} = 2.2\%$, $\delta{\Omega_bh^2} = 7.8\%$, $\delta{n_s} = 9.5\%$, $\delta{w_0} = 14\%$, $\delta{w_a} = 34\%$, $\delta{\Omega_ch^2} = 44\%$, and $\delta{b_{\textrm{\hi\,}}} = 176\%$.

 \subsubsection{Effect of foreground residuals}
 \label{sec:foregrounds}


Our results so far have considered a perfect foreground cleaning process. However, as discussed in Sect.~\ref{sec:foregorund1}, these procedures leave residuals that can bias the parameter estimation and increase the statistical noise. In order to take these effects into account, we assume that there has already been some sort of foreground removal technique and model the residual contamination as the sum of Gaussian processes with angular power spectra given by \citep{Santos:2004ju,Bull:2014rha}
\begin{equation}\label{eq:cl_fg}
C_\ell^{\rm FG}(\nu_1,\nu_2)= \epsilon^2_{\rm FG} \sum_{i}A_{i}\,\left(\frac{\ell_{\rm ref}}{\ell}\right)^{\beta_{i}}\,
\left(\frac{\nu_{\rm ref}^2}{\nu_1\,\nu_2}\right)^{\alpha_{i}}
\exp\left(-\frac{\log^2(\nu_1/\nu_2)}{2\,\xi_{i}^2}\right)\,.
\end{equation}
We assume four foreground contributions with the parameters described in Table~\ref{tab:FG}. The overall scaling, $\epsilon_{\rm{FG}}$, parameterizes the efficiency of the foreground removal technique, with $\epsilon_{\rm{FG}} = 1$ corresponding to no foreground removal.
\begin{table}
\begin{center}
\begin{tabular}{|c|c|c|c|c|}
\hline
Foreground              & A (mK$^2$) & $\beta$ & $\alpha$ & $\xi$ \\
\hline
Galactic synchrotron    & 700        & 2.4 & 2.80 & 4.0 \\
Point sources           &  57        & 1.1 & 2.07 & 1.0 \\
Galactic free-free      & 0.088      & 3.0 & 2.15 & 35  \\
Extragalactic free-free & 0.014      & 1.0 & 2.10 & 35  \\
\hline
\end{tabular}
\end{center}
\caption{Foreground model parameters taken from \citet{Santos:2004ju} with $\ell_{\rm ref}=1000$ and $\nu_{\rm ref}=130\,{\rm MHz}$.}
\label{tab:FG}
\end{table}

\begin{figure}[h!]
\centering
\includegraphics[scale=0.6]{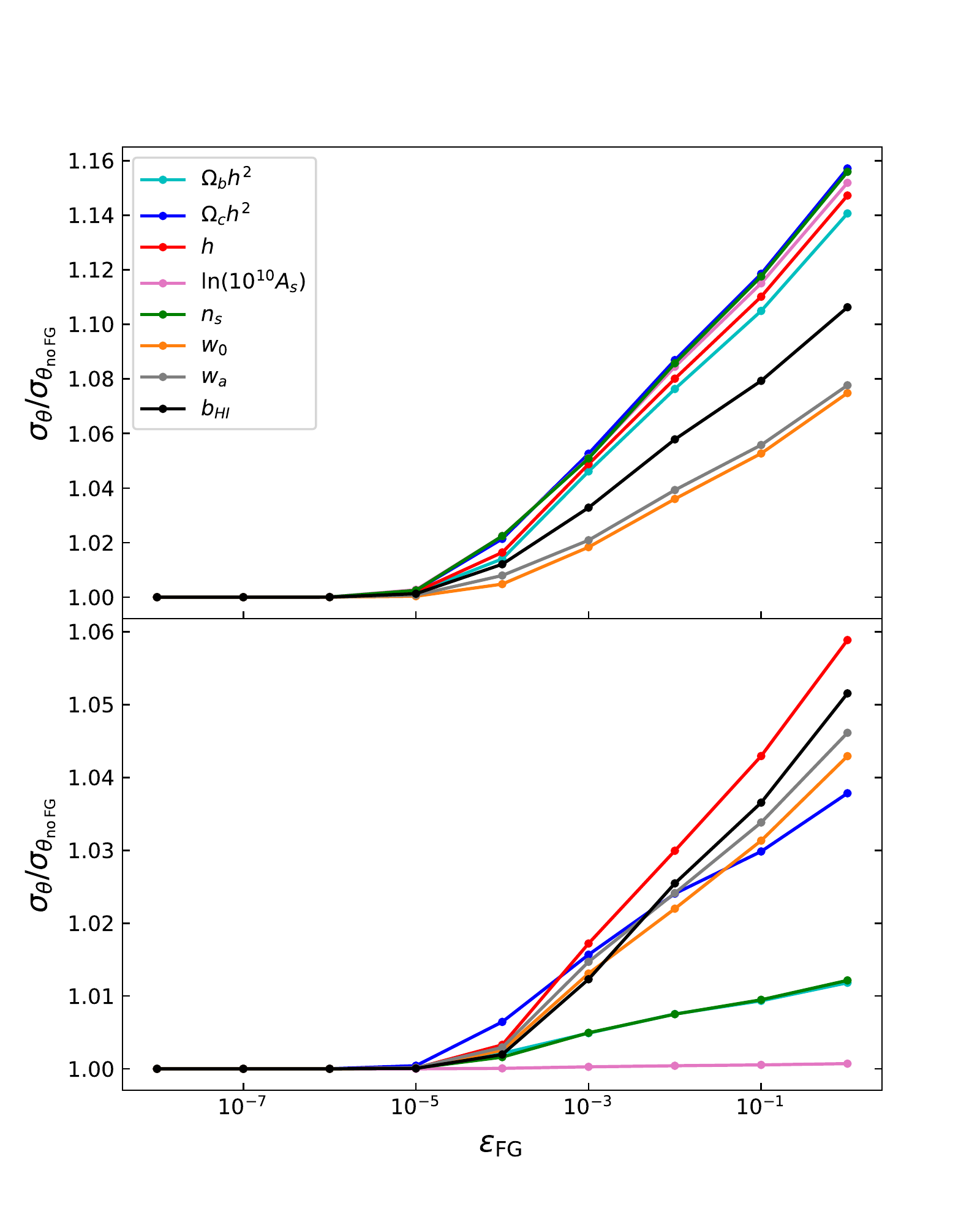}
\caption{\label{fig:fg_cov_nobias} Marginalized statistical errors as a function of the residual foreground contamination amplitude, $\epsilon_{\rm{FG}}$, from BINGO (top) and BINGO + \textit{Planck} (bottom).}
\end{figure}

The additional contribution from foreground residuals increases the covariance given by Eq.~(\ref{eq:cov}), and consequently enlarges the statistical errors on the final cosmological parameters. {The ratios of the marginalized $1\sigma$ constraints with respect to the case without foreground residuals are plotted} in Fig.~\ref{fig:fg_cov_nobias} as a function of the residual amplitude $\epsilon_{\rm{FG}}$. We find the constraints are degraded by at most $16\%$ in the case of BINGO only, and $6\%$ for BINGO + \textit{Planck}. If the efficiency of foreground removal is $\epsilon_{\rm{FG}} < 10^{-5}$, these effects are negligible.

\begin{figure}[h!]
\centering
\includegraphics[scale=0.6]{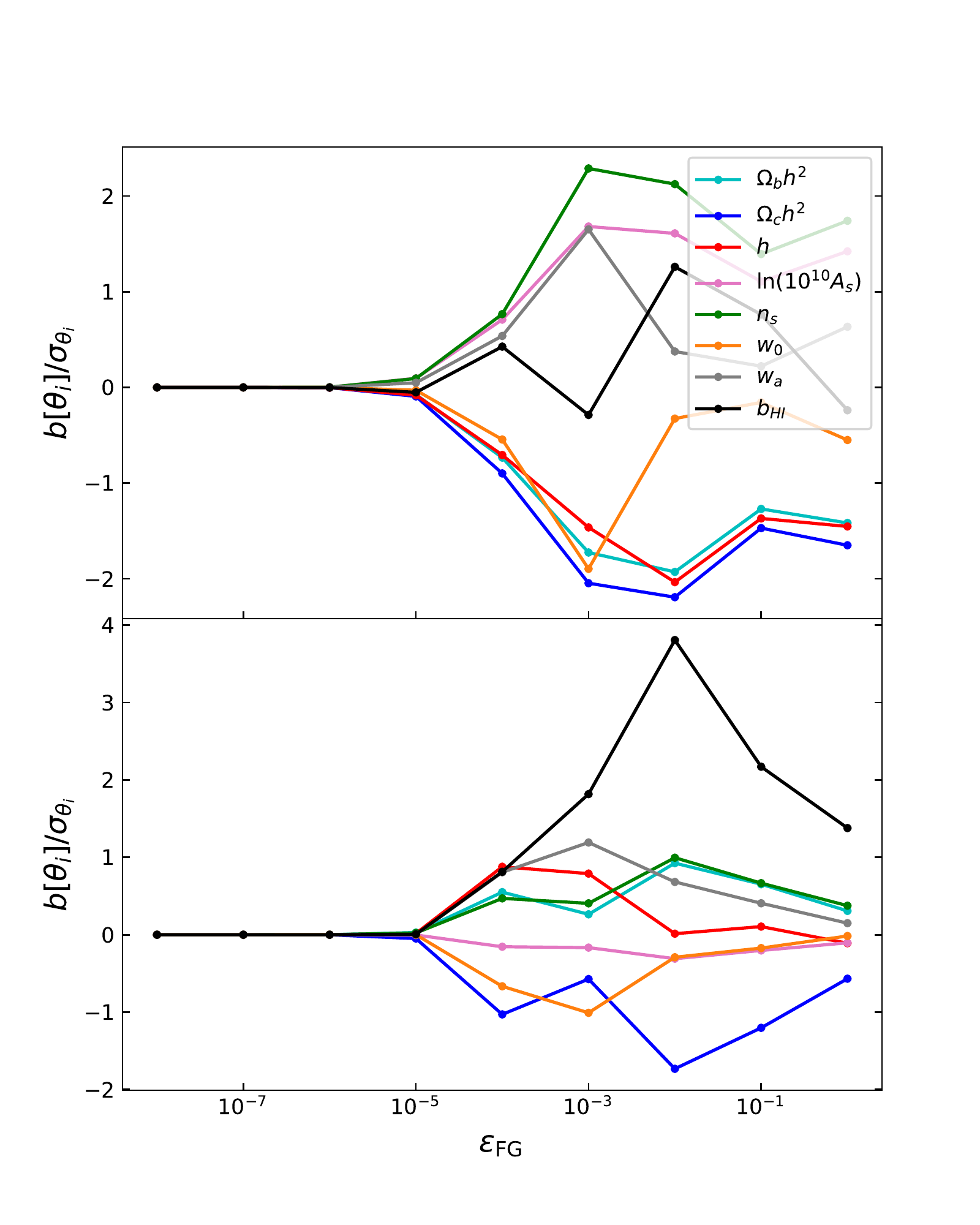}
\caption{\label{fig:fg_bias_sigma} Ratio of bias to the marginalized statistical errors for each parameter as a function of the residual foreground contamination amplitude, $\epsilon_{\rm{FG}}$, from BINGO (top) and BINGO + \textit{Planck} (bottom).}
\end{figure}

\begin{figure}[h!]
\centering
\includegraphics[scale=0.6]{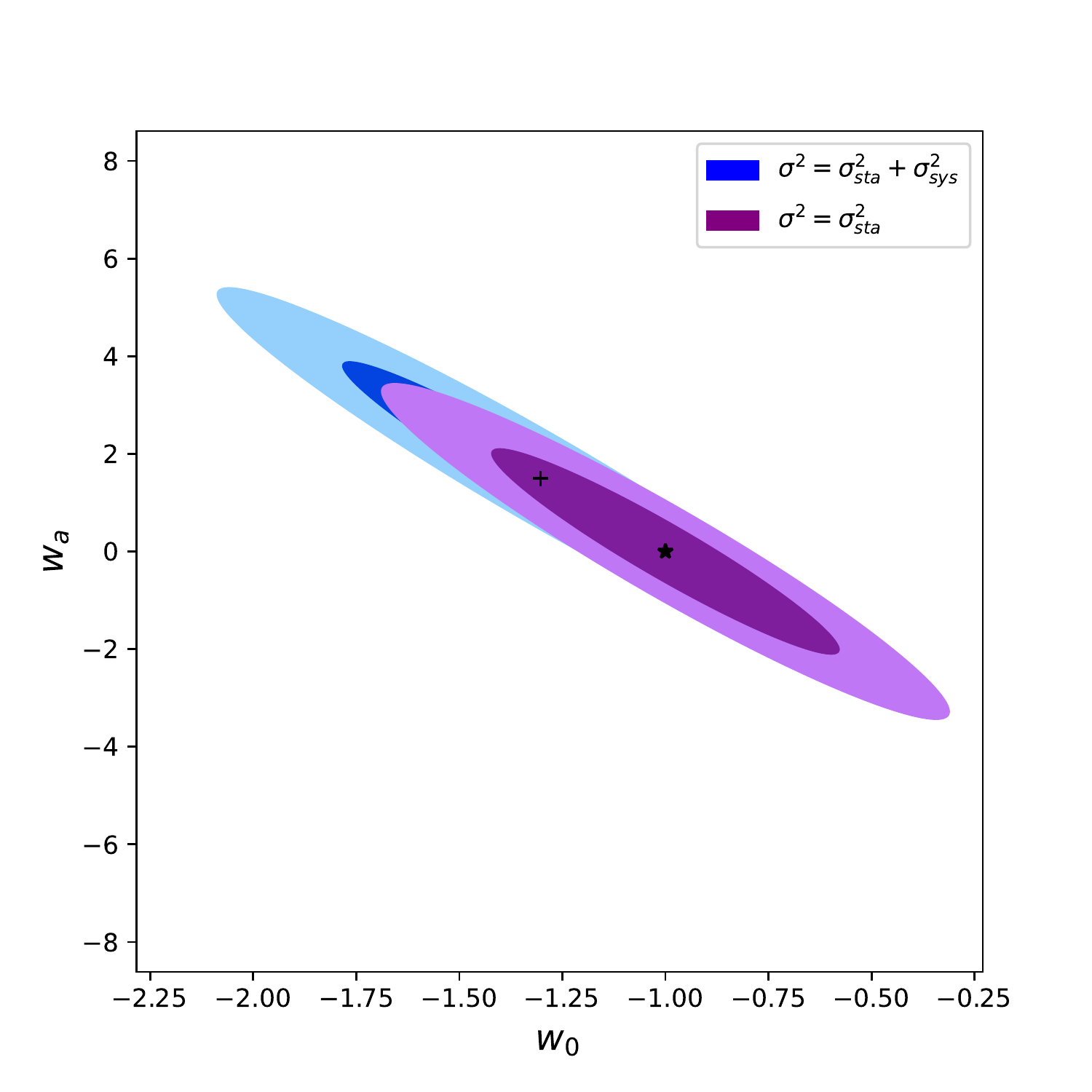}
\caption{\label{fig:fg_waxw0}Marginalized constraints (68\% and 95\% CL) on the DE EoS parameters for the CPL parameterization with BINGO, considering the effect of foreground residuals on the statistical and systematic errors. The star corresponds to our fiducial value and the cross shows the value recovered given the Fisher matrix bias. We have fixed $\epsilon_{\rm{FG}} = 10^{-4}$.}
\end{figure}

The residuals will also shift the cosmological parameters from their true value according to Eq.~(\ref{eq:fisher_bias}), with $C_\ell^{\textrm{sys}} = C_\ell^{\textrm{FG}}$. Figure~\ref{fig:fg_bias_sigma} shows the ratio of bias to the marginalized statistical errors as a function of $\epsilon_{\rm{FG}}$. We find that unlike the marginalized constraints that always decrease as $\epsilon_{\rm{FG}}$ goes to zero, the bias presents an unpredictable behavior for large values of the residual amplitude. This can be understood from Eq.~(\ref{eq:Bj}), which basically depends on the foreground residuals as $\frac{C_\ell^{\rm{FG}}}{(C_\ell + C_\ell^{\rm{shot}} + N_\ell + C_\ell^{\rm{FG}})^2}$. This function increases as we decrease $C_\ell^{\rm{FG}}$, and has a maximum at $C_\ell^{\rm{FG}} = C_\ell + C_\ell^{\rm{shot}} + N_\ell$. The bias procedure in Eq.~(\ref{eq:fisher_bias}) is valid when $C_\ell^{\rm{sys}}$ is small  compared to ${\bf C}$, which in our case happens for $\epsilon_{\rm{FG}} \lesssim 10^{-4}$. Considering a foreground removal process with $\epsilon_{\rm{FG}} = 10^{-4}$, the largest bias was $|b[\Omega_ch^2]| = 0.9\sigma$ and $|b[\Omega_ch^2]| = 1\sigma$ for BINGO and BINGO + \textit{Planck}, respectively.
Figure~\ref{fig:fg_waxw0} shows the marginalized constraints for the DE EoS parameters considering both statistical and systematic errors with BINGO.

\subsubsection{Comparison with SKA}\label{sec:ska}
\begin{figure*}
   \resizebox{\hsize}{!}
            {\includegraphics{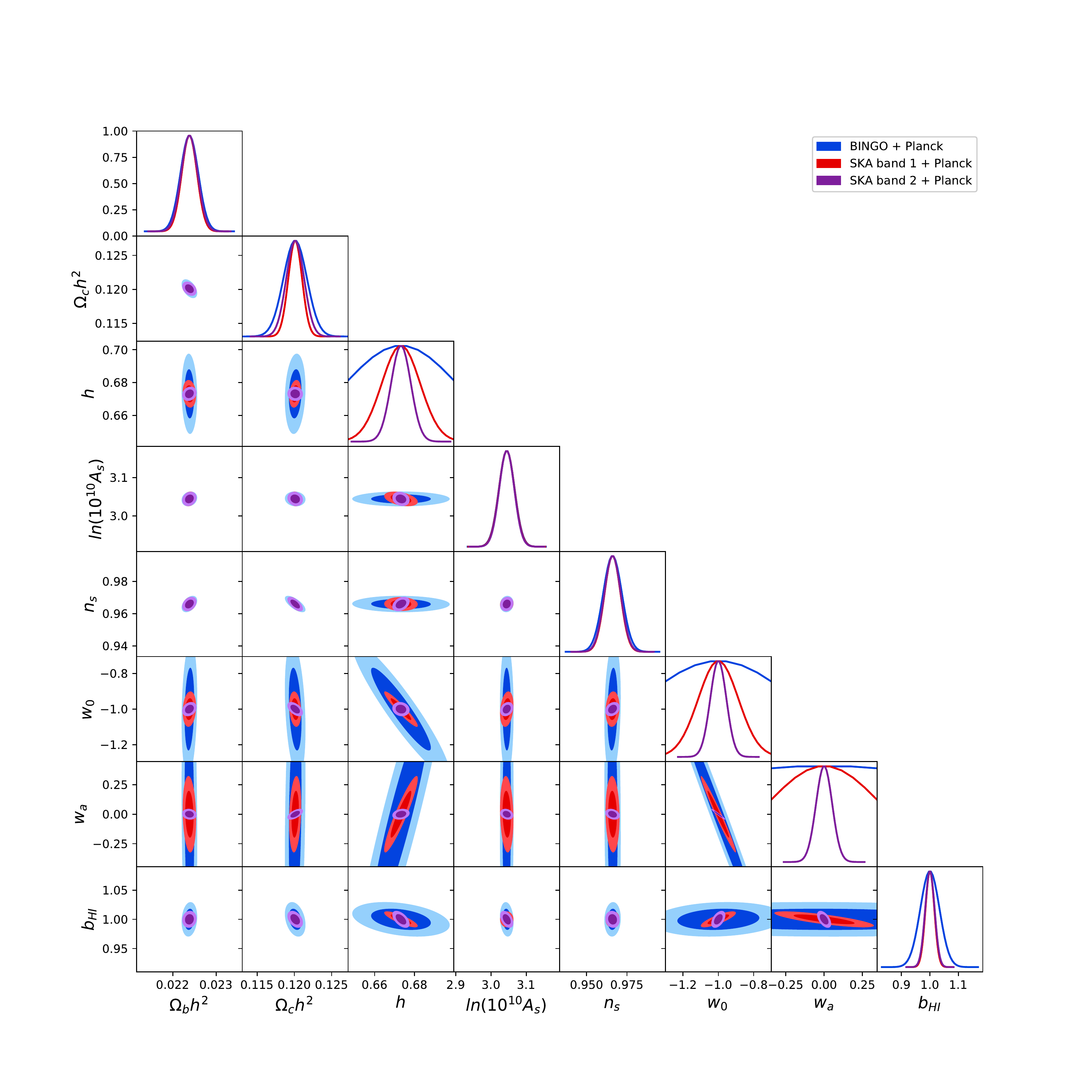}}
      \caption{One- and two-dimensional ($68\%$ and $95\%$ CL) cosmological constraints for a CPL parameterization from BINGO, SKA1-MID band 1, and SKA1-MID band 2 in combination with \textit{Planck} data. This illustrates that BINGO can be considered  a pathfinder for constraints obtained by SKA.}
      \label{plot_bingo_ska}
\end{figure*}

We now  compare the expected constraints from BINGO with the experiment design for SKA1-MID band 1 and SKA1-MID band 2 \citep{Bacon:2018dui}. We consider the same experimental setup for SKA as used in \cite{Chen:2019jms}, except that we  use a bandwidth of $10$ MHz. In order to have a proper comparison between them, in this section we only use   28 redshift bins for BINGO, which implies a bandwidth of 10 MHz, consistent with the value used for the SKA results. This   produces a small degradation in our forecast in comparison with the fiducial scenario presented in Table~\ref{tab:CPL}. We compare the marginalized constraints on our {eight-parameter space} in Fig.~\ref{plot_bingo_ska} and Table~\ref{tab:bingo_ska}. For a better comparison, we also repeat the constraints from {\it Planck} alone from  Table~\ref{tab:CPL}.

\begin{table*}
\footnotesize
\caption{\label{tab:bingo_ska} Expected $1\sigma$ constraints on the CPL cosmological parameters from BINGO, SKA1-MID band 1, and SKA1-MID band 2 in combination with \textit{Planck}.}
\begin{center}

\sisetup{
        round-mode=figures,
        round-precision=2,
        scientific-notation = fixed,
        fixed-exponent = 0
}

\begin{tabular}{|c|S|S|S|S|}
\cline{2-5}
\multicolumn{1}{c}{}  &
\multicolumn{1}{|c|}{\textit{Planck}} & 
\multicolumn{1}{|c|}{BINGO + \textit{Planck}} & 
\multicolumn{1}{|c|}{SKA Band 1 + \textit{Planck}} & 
\multicolumn{1}{|c|}{SKA Band 2 + \textit{Planck}} \\
\hline
 
\multicolumn{1}{|c|}{Parameter} &
\multicolumn{1}{|c|}{$\pm 1\sigma$ ($100\% \times\sigma/\theta_i^{\rm{fid}}$)}  &
\multicolumn{1}{|c|}{$\pm 1\sigma$ ($100\% \times\sigma/\theta_i^{\rm{fid}}$)}  &
\multicolumn{1}{|c|}{$\pm 1\sigma$ ($100\% \times\sigma/\theta_i^{\rm{fid}}$)}  &
\multicolumn{1}{|c|}{$\pm 1\sigma$ ($100\% \times\sigma/\theta_i^{\rm{fid}}$)}  \\
 
 \hline
$\Omega_b h^2$ & 0.00016 \textrm{  (0.7\%)} & 1.45606809e-04 \textrm{  (0.7\%)} & 1.25644498e-04 \textrm{  (0.6\%)} & 0.00013435 \textrm{  (0.6\%)} \\
$\Omega_c h^2$ & 0.0013 \textrm{  (1.1\%)} & 1.11645534e-03 \textrm{  (0.9\%)}  & 6.52540007e-04 \textrm{  (0.5\%)} & 0.0008447 \textrm{  (0.7\%)} \\
$h$            & 0.088 \textrm{  (13\%)} & 1.96953704e-02 \textrm{  (2.9\%)}  & 6.79397545e-03 \textrm{  (1\%)} & 0.00350296 \textrm{  (0.5\%)}  \\
${\rm{ln}}(10^{10}A_s)$ & 0.016 \textrm{  (0.5\%)} & 1.56867321e-02 \textrm{  (0.5\%)} & 1.54071935e-02 \textrm{  (0.5\%)} & 0.01506835 \textrm{  (0.5\%)} \\
$n_s$          & 0.0044 \textrm{  (0.5\%)} & 4.10272063e-03 \textrm{  (0.4\%)}  & 3.50165066e-03 \textrm{  (0.4\%)} & 0.00360917 \textrm{  (0.4\%)} \\
$w_0$          & 0.46 \textrm{  (46\%)} & 3.05375268e-01 \textrm{  (31\%)}  & 8.01957280e-02 \textrm{  (8\%)} & 0.03230461 \textrm{  (3.2\%)}  \\
$w_a$          & 1.8 & 1.22388700e+00  & 2.61905003e-01 & 0.03729145 \\
$b_{\textrm{\hi\,}}$       &  & 2.38386683e-02 \textrm{  (2.4\%)} & 1.09860341e-02 \textrm{  (1.1\%)} & 0.01172231 \textrm{  (1.2\%)} \\
 \hline
\end{tabular}
\end{center}

\end{table*}

Even considering the power of SKA, we can observe that several cosmological parameters are mainly constrained by {\it Planck}, although small improvements are still possible, especially breaking degeneracies in the parameter space. Their main contributions are in the DE EoS parameters and the Hubble constant. The Hubble constant  improved from a projected constraint of $13\%$ with {\it Planck} alone to $2.9\%$, $1\%$, and $0.5\%$ in combination with BINGO, SKA Band 1, and SKA Band 2, respectively. In addition, the DE EoS parameter $w_0$ goes from $46\%$ to $31\%$, $8\%$, and $3.2\%$. Finally, we  obtained a projected constraint for $w_a$ of $1.8$, $1.2$, $0.26$, and $0.037$ from {\it Planck}, BINGO + {\it Planck}, SKA Band 1 + {\it Planck}, and SKA Band 2 + {\it Planck}, respectively. For all these parameters, SKA Band 2 showed the best constraints. Generally, these results are in agreement with what was found in \cite{Chen:2019jms} for the SKA. Some discrepancies may be related to our larger number of bins and the fact that we are considering dimensional $C_\ell$s. This better takes into account the dependence with the brightness temperature in the Fisher matrix derivatives.

Both SKA band 1 and band 2 will survey a larger fraction of the sky than BINGO. They are also designed to explore a wider redshift range. In addition, although BINGO has a better angular resolution, the number of antennas is much lower than   for SKA and the system temperature is higher. Combining all these aspects favors SKA in the ability to constrain cosmology. On the other hand, SKA will require a more complicated technique {combining the dish array for} a IM single-dish mode. Therefore, the simplicity of the BINGO instrument will be a pathfinder for IM with SKA. In particular, Table~\ref{tab:bingo_ska} shows BINGO can provide valuable information for the \hi\, bias.

\subsection{\hi\,density and bias}\label{sec:HI_density_bias}
Using the 21cm line of \hi\, as a tracer of the underlying matter distribution requires the knowledge of the \hi\, mean density and bias. If we are only interested in the cosmological constraints, they can be considered  nuisance parameters that we marginalize over. On the other hand, we can also use our 21cm survey to learn about their distribution and evolution.

In the previous sections we  fix the value for the \hi\, density parameter, $\Omega_{\textrm{\hi\,}} = \rho_{\textrm{\hi\,}}/\rho_{c}$, and described the bias by a constant in the whole redshift range. We  now extend these assumptions and see their impact in our cosmological parameters and our ability to constrain them with BINGO.

First we fix the \hi\, density parameter and assume a constant bias. Table~\ref{tab:LCDM} shows that we can achieve a $4.1\%$ precision measurement in the bias with BINGO under the $\Lambda$CDM model, which can be  improved  to $1.1\%$ if we combine with \textit{Planck} data. Beyond the $\Lambda$CDM model, the constraints are degraded by the extra cosmological parameters, but we still obtain a $2.3\%$ precision measurement at $1\sigma$ in the joint analysis BINGO + \textit{Planck}, as calculated in Tables~\ref{tab:wCDM} and \ref{tab:CPL}.

Next, we extend this simplest model and allow $\Omega_{\textrm{\hi\,}}$ to be a free parameter. Equations~(\ref{eq:Tb_mean}), (\ref{eq:cl}), and (\ref{eq:DW})   tell us that in this case $\Omega_{\textrm{\hi\,}}$ and $A_s$ are completely degenerate, and therefore we cannot constrain them with BINGO alone. In order to break that degeneracy, we combine our analysis with \textit{Planck}. Our results are presented in Table~\ref{tab:HI_bias}. We obtain
\begin{align}
    \sigma_{\Omega_{\textrm{\hi\,}}} & = 4.1\times10^{-5} \quad (6.6\%) \,, \\
    \sigma_{b_{\textrm{\hi\,}}} & = 3.6\times10^{-2} \quad (3.6\%) \,.
\end{align}
\begin{table*}
\footnotesize
\caption{\label{tab:HI_bias} Expected $1\sigma$ constraints on the CPL cosmological parameters from BINGO + \textit{Planck} for three different models of the \hi\, density and bias.}
\begin{center}

\sisetup{
        round-mode=figures,
        round-precision=2,
        table-format = 1.2e1,
        table-space-text-post = \textrm((00.00\%))
}

\begin{tabular}{|c|S[table-number-alignment = left,table-align-text-post = false]|S[table-number-alignment = left,table-align-text-post = false]|S[table-number-alignment = left,table-align-text-post = false]|}
\cline{2-4}
\multicolumn{1}{c}{}  &
\multicolumn{1}{|c|}{$\Omega_{\textrm{\hi\,}} = \text{const.}$} & 
\multicolumn{1}{|c|}{$\Omega_{\textrm{\hi\,}}(z)$} & 
\multicolumn{1}{|c|}{$b_{\textrm{\hi\,}}(z)$} \\
\hline
 
\multicolumn{1}{|c|}{Parameter} &
\multicolumn{1}{|c|}{$\pm 1\sigma$ ($100\% \times\sigma/\theta_i^{\rm{fid}}$)}  &
\multicolumn{1}{|c|}{$\pm 1\sigma$ ($100\% \times\sigma/\theta_i^{\rm{fid}}$)}  &
\multicolumn{1}{|c|}{$\pm 1\sigma$ ($100\% \times\sigma/\theta_i^{\rm{fid}}$)}  \\
 
 \hline
$\Omega_b h^2$ & 1.50496736e-04 \textrm{  (0.7\%)} & 1.51226976e-04 \textrm{  (0.7\%)} & 1.51238660e-04 \textrm{  (0.7\%)} \\
$\Omega_c h^2$ & 1.25778640e-03 \textrm{  (1\%)} & 1.27321329e-03 \textrm{  (1.1\%)} & 1.27393186e-03 \textrm{  (1.1\%)} \\
$h$            & 1.95432615e-02 \textrm{  (2.9\%)}  & 2.35851220e-02 \textrm{  (3.5\%)} & 2.41306278e-02 \textrm{  (3.6\%)}  \\
${\rm{ln}}(10^{10}A_s)$ & 1.57706899e-02 \textrm{  (0.5\%)} & 1.57708390e-02 \textrm{  (0.5\%)} & 1.57708707e-02 \textrm{  (0.5\%)} \\
$n_s$          & 4.29295655e-03 \textrm{  (0.4\%)}  & 4.32043598e-03 \textrm{  (0.4\%)} & 4.32051013e-03 \textrm{  (0.4\%)} \\
$w_0$          & 3.07534883e-01 \textrm{  (31\%)}  & 3.27912310e-01 \textrm{(  33\%)} & 3.43884651e-01 \textrm{  (34\%)}  \\
$w_a$          & 1.31551925e+00  & 1.38510189e+00  & 1.46136966e+00  \\
$b_{\textrm{\hi\,}}$ or $b_{\textrm{\hi\,}}^1$       & 3.63570429e-02 \textrm{  (3.6\%)} & 3.65841587e-02 \textrm{  (3.7\%)} & 6.87464645e-02 \textrm{  (6.9\%)} \\
$b_{\textrm{\hi\,}}^2$       &   &   & 5.76025558e-02 \textrm{  (5.8\%)} \\
$b_{\textrm{\hi\,}}^3$       &   &   & 5.88316096e-02 \textrm{  (5.9\%)} \\
$\Omega_{\textrm{\hi\,}}$ or $\Omega_{\textrm{\hi\,}}^1$  & 4.10098678e-05  \textrm{  (6.6\%)} & 4.47919239e-05 \textrm{  (7.2\%)} & 5.41998486e-05 \textrm{  (8.7\%)} \\
$\Omega_{\textrm{\hi\,}}^2$       &   & 4.61174034e-05 \textrm{  (7.4\%)} & 5.19442019e-05 \textrm{  (8.4\%)} \\
$\Omega_{\textrm{\hi\,}}^3$       &   & 4.79398301e-05 \textrm{  (7.7\%)} & 5.60110668e-05 \textrm{  (9\%)} \\
 \hline
\end{tabular}
\end{center}

\end{table*}
Compared with the fiducial model presented in Table~\ref{tab:CPL}, the parameter most affected by the inclusion of $\Omega_{\textrm{\hi\,}}$ is the bias, with a degradation of $61\%$. The next most sensitive is the DM density parameter, $\Omega_ch^2$, which is degraded by $15\%$, followed by the DE EoS parameter, $w_a$, which has its uncertainty increased by $9.5\%$. The other parameters change by at most $\sim 5\%$.

A natural extension to this model is to assume that the \hi\, parameters evolve with redshift. 
We  use a stepwise model, which is simple and provides an agnostic way to describe the evolution of these parameters over redshift. Therefore, we  combine our 30 redshift bins into three groups and define the parameters as $\Omega_{\textrm{\hi\,}}^i$ and $b_{\textrm{\hi\,}}^i$ inside each group. We first consider the \hi\, density as a function of redshift, keeping the bias constant. In this case our constraints are given by
\begin{align}
    \sigma_{\Omega_{\textrm{\hi\,}}^1} & = 4.5\times10^{-5} \quad (7.2\%) \,, \\
    \sigma_{\Omega_{\textrm{\hi\,}}^2} & = 4.6\times10^{-5} \quad (7.4\%) \,, \\
    \sigma_{\Omega_{\textrm{\hi\,}}^3} & = 4.8\times10^{-5} \quad (7.7\%) \,, \\
    \sigma_{b_{\textrm{\hi\,}}} & = 3.7\times10^{-2} \quad (3.7\%) \,,
\end{align}
where the indices 1, 2, and 3 represent each group of ten redshift bins. The projected constraints are very similar in all three groups of bins, but lower redshifts  show slightly better results. In addition, the projected bias is only marginally affected by the extra degrees of freedom in the \hi\, density parameter.

Finally, we consider the case where both $\Omega_{\textrm{\hi\,}}$ and $b_{\textrm{\hi\,}}$ are functions of redshift. Our results are presented in Table~\ref{tab:HI_bias} together with the previous scenarios. The extra parameters describing the \hi\, distribution and evolution  mainly affect our uncertainties on the cosmological parameters describing the DE EoS (degrading by $\delta w_0 = 4.9\%$ and $\delta w_a = 5.5\%$) and the Hubble constant (with $\delta h = 2.3\%$) compared with the previous case. The other CPL parameters are not significantly impacted as \textit{Planck} is  mainly responsible for their constraints. On the other hand, the \hi\, bias constraints vary by at most $88\%$. This last scenario can put constraints on the \hi\, parameters with uncertainties of around $8.5\%$ and $6\%$ for the \hi\, density parameter and bias, respectively:
\begin{align}
    \sigma_{\Omega_{\textrm{\hi\,}}^1} & = 5.4\times10^{-5} \quad (8.7\%) \,, \\
    \sigma_{\Omega_{\textrm{\hi\,}}^2} & = 5.2\times10^{-5} \quad (8.4\%) \,, \\
    \sigma_{\Omega_{\textrm{\hi\,}}^3} & = 5.6\times10^{-5} \quad (9\%) \,, \\
    \sigma_{b_{\textrm{\hi\,}}^1} & = 6.9\times10^{-2} \quad (6.9\%) \,, \\
    \sigma_{b_{\textrm{\hi\,}}^2} & = 5.8\times10^{-2} \quad (5.8\%) \,, \\
    \sigma_{b_{\textrm{\hi\,}}^3} & = 5.9\times10^{-2} \quad (5.9\%) \,.
\end{align}

Further details about the \hi\, distribution using N-body simulations can be obtained in our BINGO companion paper \citep{2020_mock_simulations}.
\subsection{Massive neutrinos}\label{sec:massive_neutrinos}

It is well known that neutrinos should have a low mass in order to explain their change of flavors, observed in both solar and atmospheric neutrinos \citep{Fukuda:1998mi,Ahmad:2002jz}. However, the experiments only allow us to determine two squared mass differences. On the other hand, it is possible to constrain the sum of neutrino masses, $\sum m_\nu$, from a combination of the CMB and matter power spectrum.

Massive neutrinos can impact the CMB spectrum in different ways. At the background level they may change the redshift of matter-to-radiation equality, the angular diameter distance to the last scattering surface, and the late ISW effect, while neutrino perturbations affect the early ISW effect \citep{Lesgourgues:2012uu}. In order to analyze the constraints on the total neutrino mass, we consider an extension to the $\Lambda$CDM model allowing for an extra degree of freedom on the sum of neutrino masses, $\Lambda$CDM + $\sum m_{\nu}$. {We  performed a MCMC sampling} with the \textit{Planck} 2018 TT + TE + EE + lowE likelihood, assuming one massive neutrino with total mass equal to $\sum m_{\nu}$. Our result is given by
\begin{equation}
    \sigma_{\sum m_{\nu}} < 0.36 \, \textrm{eV} \quad (95\% \, \textrm{CL} \,, \textit{Planck}) \,.
\end{equation}
This is higher than the value of $0.26 \, \textrm{eV}$ reported in \cite{Aghanim:2018eyx} using the \texttt{Plik} likelihood, but below the value of $0.38 \, \textrm{eV}$ from the \texttt{CamSpec} likelihood. For the purpose of this forecast paper, we  use the value we report above.

The combination of CMB data with other cosmological observations, such as measurements of the \hi\, power spectrum, can break the geometric degeneracy in the parameter space of the $\Lambda$CDM + $\sum m_{\nu}$ model and improve the constraints. We  combined the covariance matrix from the \textit{Planck} data with our 21cm Fisher analysis, and this leads to
\begin{equation}
    \sigma_{\sum m_{\nu}} < 0.14 \, \textrm{eV} \quad (95\% \, \textrm{CL} \,, \textrm{BINGO} + \textit{Planck}) \,,
\end{equation}
which is slightly higher than what was obtained in \cite{Aghanim:2018eyx} for the combination between \textit{Planck} temperature and polarization with other BAO data. Therefore, if BINGO's systematic noise can be properly taken into account, it will be able to put competitive constraints on the sum of neutrino masses compared with the present available constraints, but using a completely independent tracer of LSS.

\subsection{Alternative cosmologies}\label{sec:alt_cosmo}

\subsubsection{Modified Gravity: $\mu$ and $\Sigma$ parameterization}

An alternative explanation for the accelerated expansion of the Universe is to invoke modifications of GR. To properly describe the current data, these modifications must explain the current acceleration with background dynamics very close to the $\Lambda$CDM predictions today, while maintaining the local and astrophysical tests of GR. There are several proposed modifications of gravity in the literature  \citep[for a review see][]{Clifton:2011jh} with the most studied being scalar-tensor theories and higher-derivative theories, such as $f(R)$. The $f(R)$ theory is the simplest modification of GR and can be mapped  onto a scalar-tensor theory via field redefinition and conformal transformation.  A more general class of scalar-tensor theories are the Horndeski models \citep{Horndeski:1974wa}, a class of models that modify GR while still maintaining up to second-order derivatives in the equations of motion, avoiding instabilities \citep{Deffayet:2011gz}. Although the background evolution of these modifications of GR needs to behave close to $\Lambda$CDM today, the evolution of their perturbations might behave very differently, and that represents an avenue to test these models.
        
            

{Modified gravity (MG)} models can be parameterized through two phenomenological functions, $\mu$ and $\gamma$. On sub-Hubble scales, the Poisson equation and the relation between the gravitational potentials receive corrections if gravity is modified, given by
\begin{align}
    -k^2 \Psi & = 4 \pi G a^2 \mu(a,k) \bar{\rho} \Delta\,, \nonumber \\
    \Phi & = \gamma(a,k) \Psi\,,
\label{eq:potentials_perturbation}
\end{align}
where $\bar{\rho}$ is the background value of the matter energy density, $\Delta=\delta+3 \mathcal{H} v/k$ is the comoving density perturbation with $\mathcal{H}$ the conformal time Hubble parameter, $\delta$ is the density contrast, and where the anisotropic stress from relativistic species was neglected. The parameters $\mu$ and $\gamma$ represent deviations from GR, since in GR $\mu=\gamma=1$ at all times, while for alternative models both functions can depend upon time and scale. Using these equations, the variation of the energy-momentum tensor of a scalar tensor theory in the Jordan frame yields the continuity and Euler equations on sub-Hubble scales \citep{Hall:2012wd}
\begin{align}
    &\Ddot{\delta}+\mathcal{H} \dot{\delta} - 4\pi G  a^2 \mu \bar{\rho} \delta =0\,,  \\
    & \Ddot{v}+ \left( 2 \mathcal{H} - \frac{\dot{\mu}}{\mu} \right) \dot{v}+ \left( \dot{\mathcal{H} + \mathcal{H}^2-\mathcal{H}}\frac{\dot{\mu}}{\mu} -4 \pi G a^2 \mu \bar{\rho} \right) v =0 \,.
\end{align}
These equations show that the density perturbation and velocity depend only on changes in $\mu$ on subhorizon scales. Changes in the density and velocities can be probed by experiments like BINGO since the 21cm brightness temperature is sensitive to density and RSD, offering a chance to probe the time and scale dependence of the parameter $\mu$.

To project constraints on the cosmological parameters, we use the results from BINGO combined with data from the CMB. Modifications of gravity that aim to explain the late-time acceleration only affect the CMB at perturbed level via the ISW effect in the CMB temperature anisotropies and via CMB weak lensing. This is a result of the time dependency of the potentials $\Phi$ and $\Psi$. The CMB is sensitive to modifications of the Weyl potential, given by $(\Phi + \Psi)/2$, which can be related to the density perturbation from Eq.~(\ref{eq:potentials_perturbation}),
\begin{equation}
    -k^2 (\Phi + \Psi) = 8 \pi G a^2 \Sigma (a,k) \bar{\rho} \Delta\,,
\end{equation}
where $\Sigma (a,k) \equiv \mu (1+\gamma)/2$. Due to the degeneracy between $\mu$ and $\gamma$, it can be more convenient to work with this new function $\Sigma$ \citep{Daniel:2010ky}.
            
\begin{figure}[]
\centering
\includegraphics[scale=0.6]{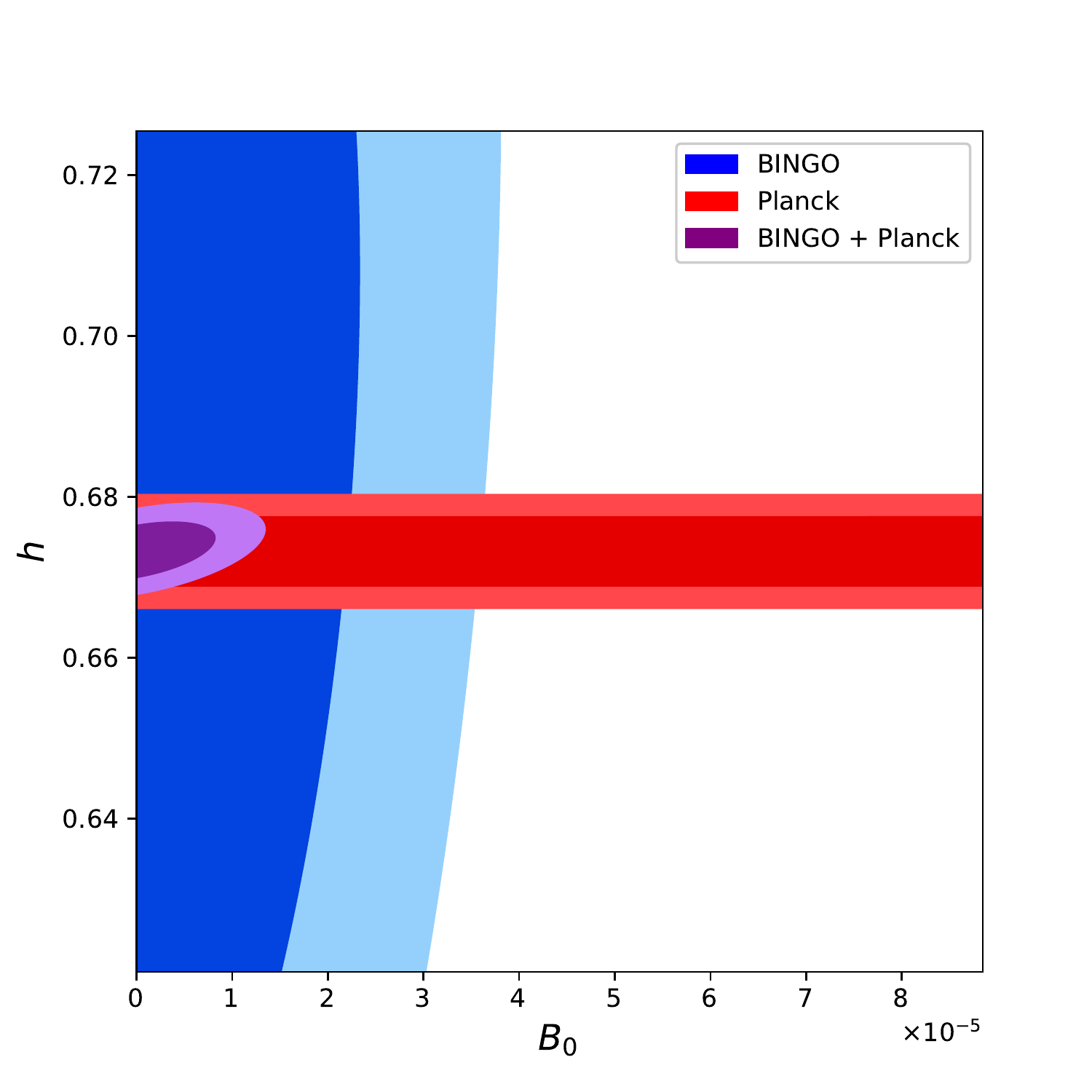}
\caption{\label{fig:mg} Marginalized constraints ($68\%$ and $95\%$) CL  on the modified gravity parameter $B_0$ and the Hubble parameter from BINGO, \textit{Planck}, and BINGO + \textit{Planck}.}
\end{figure}

In order to forecast {constraints on MG models} with the BINGO telescope, we  consider a specific form for that parameterization that is related to $f(R)$ theories of gravity. The $B_0$-parameterization of $f(R)$ gravity provides a good approximation on quasi-static scales \citep{Hu:2007nk,Giannantonio:2009gi,Hojjati:2012rf}, and has a parameterization given by
\begin{align}
    \mu(a,k) & = \frac{1}{1 - 1.4\times 10^{-8}|\lambda/{\rm Mpc}|^2a^3}\frac{1 + 4\lambda^2k^2a^4/3}{1 + \lambda^2k^2a^4}\,, \\
    \gamma(a,k) & = \frac{1 + 2\lambda^2k^2a^4/3}{1 + 4\lambda^2k^2a^4/3}\,,
\end{align}
where $B_0 \equiv 2H_0^2\lambda^2$. Therefore, considering the $\Lambda$CDM parameters plus $B_0$, we obtained a projected constraint of
\begin{align}
    \sigma_{B_0} & = 3.1\times 10^{-5}\quad \textrm{(BINGO),} \\
    \sigma_{B_0} & = 5.3\times 10^{-2}\quad \textrm{(\textit{Planck}),} \\
    \sigma_{B_0} & = 1.1\times 10^{-5}\quad \textrm{(BINGO + \textit{Planck})}\,.
\end{align}
Figure~\ref{fig:mg} shows the 2D marginalized contours for $B_0$ and $h$. We can see that BINGO will lead to a tight constraint on the MG parameter and also how the combination with \textit{Planck} breaks the degeneracy in the parameter space. We note that the tight constraint on the Hubble constant from {\it Planck} comes from the fact we have assumed the $\Lambda$CDM model for the cosmological background.

\subsubsection{Interacting dark energy}

Another possible modification of the $\Lambda$CDM is to consider an interaction in the dark sector \citep{Wetterich:1994bg,Amendola:1999er}. Since the dark sector is only detected gravitationally, different types of interaction of DE and DM are possible \citep{Wang:2016lxa}. In a field theory description of these components, this interaction is allowed and even mandatory \citep{Micheletti:2009pk}. It can also alleviate the coincidence problem, given an appropriate interaction and a dynamical mechanism to make DE leave the scaling solution and produce late acceleration \citep{Copeland:2006wr}.
        
The study of the interacting dark sector model is challenging because of the unknown nature of the two components, which makes it hard to describe the origin of such an interaction from first principles. Many different models of this interaction have been studied in the literature either from a phenomenological or from a field theory  point of view. Here we  take a purely phenomenological approach to describe the interacting dark sector (for a more general description of the classification of these models see \citealt{Wang:2016lxa}, and  \citealt{Koyama:2009gd,Pu:2014goa,Abdalla:2014cla,Costa:2013sva,Costa:2014pba,Marcondes:2016reb,Landim:2016isc,Landim:2017lyq}).
        
The interaction acts so that the energy-momentum of the dark sector components alone does not obey a conservation law,
\begin{equation}
    \nabla _{\mu} T_{(i)}^{\mu\nu} = Q_{(i)}^{\nu}\,,
\end{equation}
where conservation of the total energy-momentum implies that the right-hand sides add up to zero, $\sum _{i} Q_{(i)}^{\nu}=0$. This is how the interaction can be realized. 
        
Given the energy conservation of the full energy-momentum tensor, we represent DE and DM as fluids in a 
Friedmann-Lema\^itre-Robertson-Walker (FLRW) Universe obeying
\begin{alignat}{1}
    \dot{\rho}_{dm}+3H\rho_{dm}= & +Q\,,\nonumber \\
    \dot{\rho}_{de}+3H\left(1+w_{de}\right)\rho_{de}= & -Q\,,\label{eq:intera_fen}
\end{alignat}
where $\rho_{dm}$ and $\rho_{de}$ are the energy densities of DM and DE, respectively, and the interaction $Q$ can be a function of $\rho_{dm}$, $\rho_{de}$, or both. We assume that the EoS is constant in this simple model. By this definition if $Q>0$, we have DE decaying in DM and for $Q<0$ we have DM decaying in DE. The second case favors a Universe dominated by DM in the past and DE in the future. The interaction $Q(\rho_{dm},\rho_{de})$ can be expanded in a Taylor series, and the  phenomenological term  \citep{Feng:2008fx,He:2010im}
\begin{alignat}{1}
    Q = 3H(\xi_{dm}\rho_{dm}+\xi_{de}\rho_{de})\,
\end{alignat}
is   considered, where $\xi_{dm}$ and $\xi_{de}$ are constants.
        
This model presents two extra parameters when compared to the $\Lambda$CDM. However, due to instabilities in the DE perturbations and curvature \citep{Valiviita:2008iv,He:2008si}, the parameter space of this model is reduced;  the allowed regions are listed in Table~\ref{tab:stabilities} \citep{He:2008si,Gavela:2009cy}.  
\begin{table}[h]
    \setlength\tabcolsep{16pt} 
    \centering 
    \caption{\label{tab:stabilities} Stability conditions on the (constant) EoS and interaction for the
 interacting DE models considered in this work.}
    \renewcommand*{\arraystretch}{1.4}
    \begin{tabular}{@{}lc@{}}
        \hline 
        Case     &   Condition   \\ 
        \hline
        $Q \propto \rho_{de}$ ($\xi_{dm} = 0$) & $w < -1$ and   $\xi_{de} > 0$ ; or \\ 
        & $-1 < w < 0$ and $\xi_{de} < 0$ \\
        \hline 
      $Q \propto \rho_{dm}$ ($\xi_{de} = 0$)  & $w < -1, \, \forall \, \xi_{dm} $      \\
      \hline 
    \end{tabular}
\end{table}



In addition to the energy transfer in the background continuity equations, the phenomenological interaction will affect the time evolution of the first-order perturbations, which in the synchronous gauge are given by \citep{He:2010im,Costa:2013sva}
\begin{align}
        \dot{\delta}_{dm} & =  -\left(kv_{dm} + \frac{\dot{h}}{2}\right) + 3\mathcal{H}\xi_{de} \frac{1}{r} \left( \delta_{de}-\delta_{dm} \right)\,, \label{linear_pert_1} \\
        \dot{\delta}_{de} & = -\left(1+w \right) \left(k v_{de} + \frac{\dot{h}}{2}\right)+3\mathcal{H} (w - c_{e}^{2}) \delta_{de}  \nonumber \\
                                        & +3\mathcal{H} \xi_{dm} r \left( \delta _{de}-\delta _{dm} \right)  \nonumber \\
                                        & -3\mathcal{H} \left( c_{e}^{2}-c_{a}^{2} \right) \left[ 3 \mathcal{H} \left( 1+w \right) + 3\mathcal{H} \left( \xi_{dm} r+\xi_{de} \right) \right]\frac{v_{de}}{k}      \,, \label{linear_pert_2} \\
        \dot{v}_{dm} & = -\mathcal{H}v_{dm} -3\mathcal{H}\left(\xi_{dm} + \frac{1}{r}\xi_{de}\right)v_{dm} \,, \label{linear_pert_3} \\
        \dot{v}_{de} & = -\mathcal{H}\left(1-3c_{e}^{2} \right) v_{de}+\frac{3\mathcal{H}}{1+w} \left( 1+c_{e}^{2} \right) \left(\xi_{dm} r+\xi_{de} \right) v_{de}  \nonumber \\
                                        & + \frac{kc_{e}^{2}\delta _{de}}{1+w}\,,
        \label{eq:linear_pert_4}
\end{align}
where $\delta_{dm}$ ($\delta_{de}$) and $v_{dm}$ ($v_{de}$) refer respectively to the overdensity and peculiar velocity of DM (DE), $h$ is the metric perturbation in the synchronous gauge, $c_e$ represents the effective sound speed, $c_a$ refers to the adiabatic sound speed for the DE fluid at the rest frame, and $r \equiv \rho_{dm}/\rho_{de}$. Through Eqs.~(\ref{eq:intera_fen}) -- (\ref{eq:linear_pert_4}) the expansion history and the growth of LSS are seen to be changed by the phenomenological interaction, and thereby the deviated \hi\,  IM signals relative to standard cosmology can be used to characterize and/or constrain the interacting DE model.

We observe that Eq.~(\ref{eq:frac_Tb}) was obtained assuming the Euler equation. However, in an interacting DE model, the DM component exchanges energy-momentum with DE and does not obey the same relation as the regular matter. This can be observed in Eq.~(\ref{linear_pert_3}). This leads to an additional contribution to the fractional brightness temperature perturbation, assuming that the \hi\,  velocity follows the standard relation $v = (\bar{\rho}_bv_b + \bar{\rho}_{dm}v_{dm})/\bar{\rho}$. A more detailed discussion about the imprints that interacting DE models can leave on the 21cm power spectrum can be found in \cite{Xiao:2021nmk}.

\begin{figure}[h!]
\centering
\includegraphics[scale=0.6]{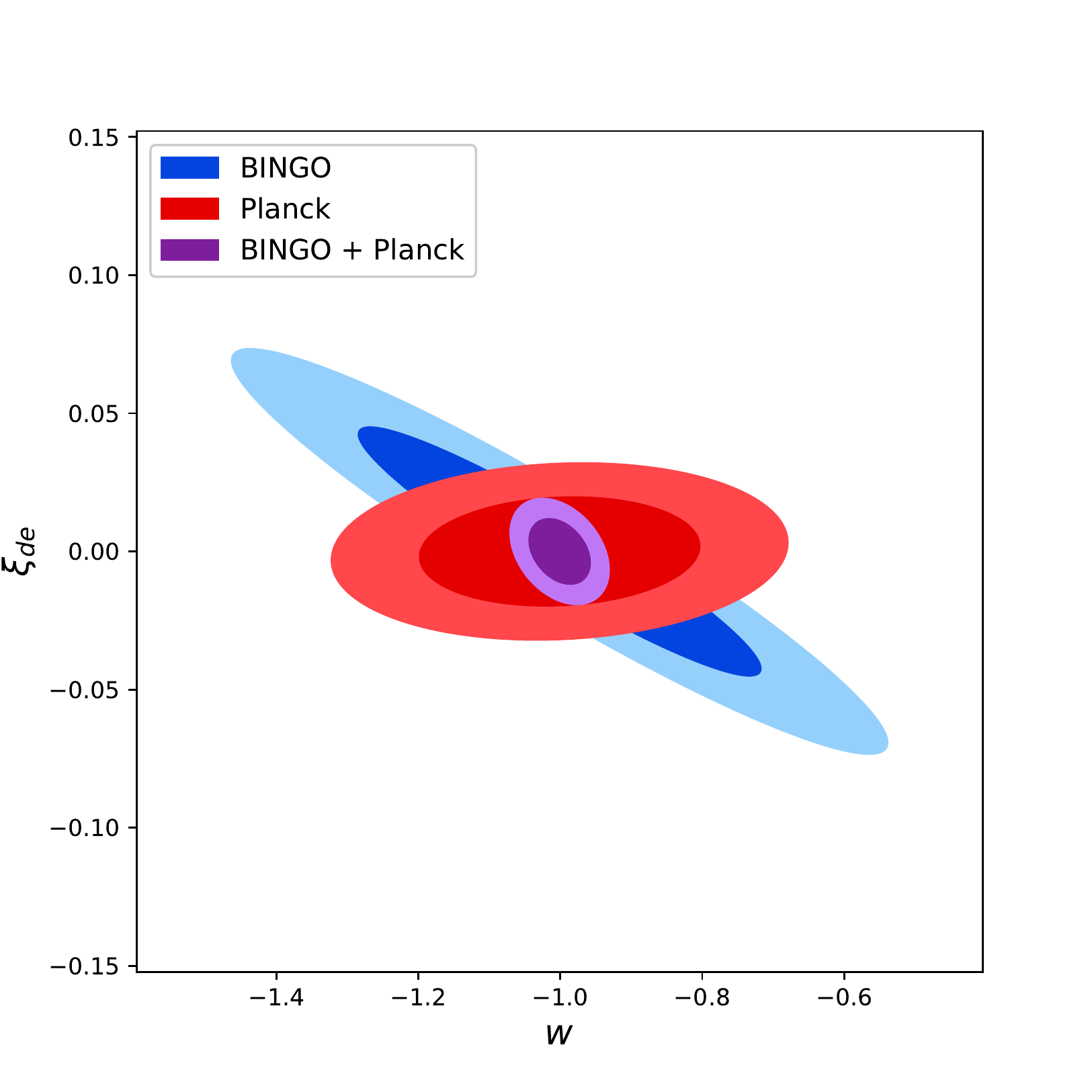}\label{plot_IDE2}
\includegraphics[scale=0.6]{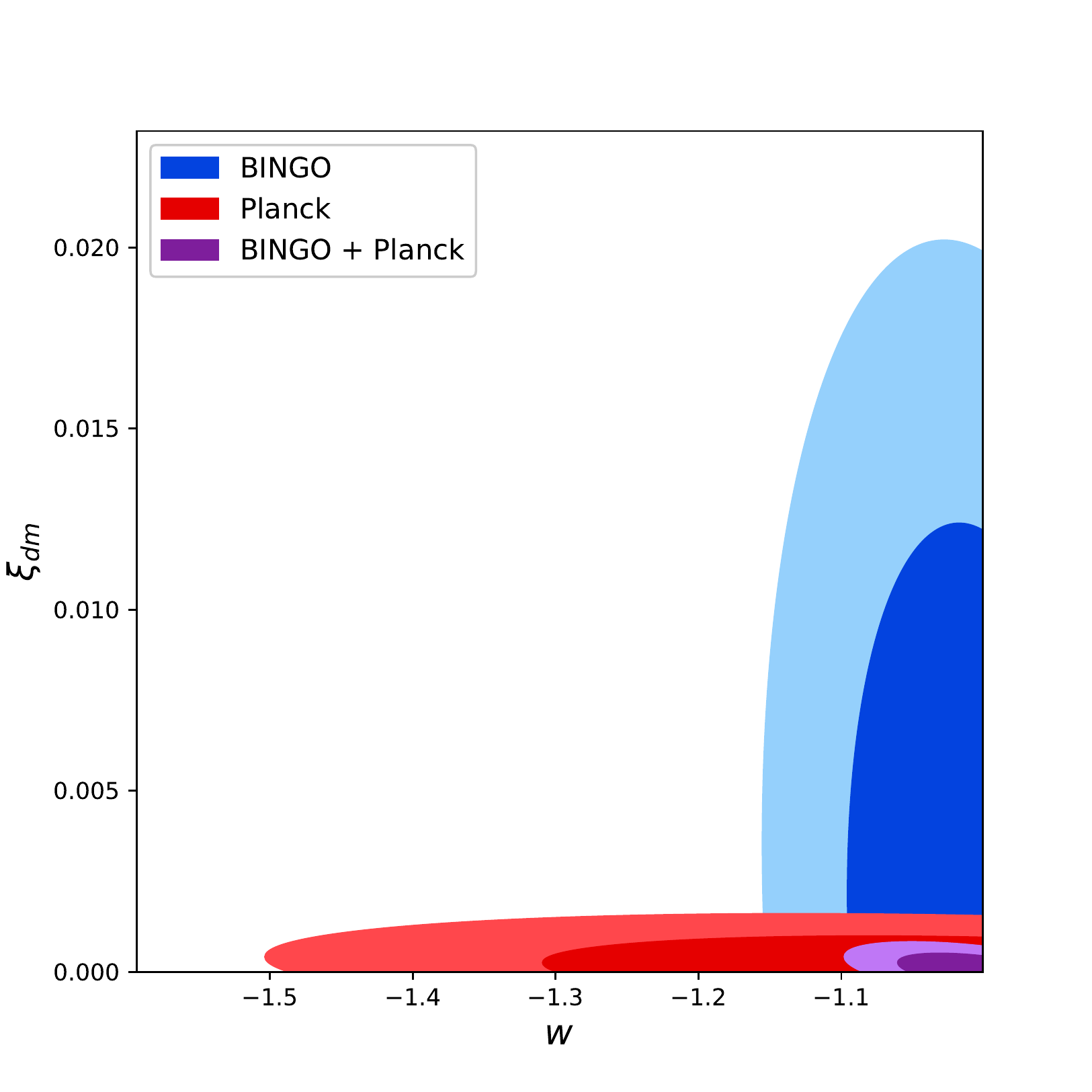}\label{plot_IDE3}
\caption{\label{plot_IDE} Marginalized constraints ($68\%$ and $95\%$ CL)  for two interacting dark energy models with $Q \propto \rho_{de}$ and $w < -1$ (top) or $Q \propto \rho_{dm}$ (bottom) from BINGO, \textit{Planck}, and BINGO + \textit{Planck}.}
\end{figure}

It is known that an interaction in the dark sector can modify the CMB spectrum at small $\ell$ and shift the acoustic peaks at large multipoles \citep{Costa:2013sva}. On the other hand, the DE EoS only modifies the low multipoles in the CMB spectrum. Combining information from the late Universe can further break degeneracies and improve the parameter constraints.  Figure~\ref{plot_IDE} presents the 2D marginalized contours for two interacting DE scenarios with $Q \propto \rho_{de}$ and $Q \propto \rho_{dm}$. Because of the stability conditions, the  $Q \propto \rho_{de}$ model needs to be divided into two regions, as described in Table~\ref{tab:stabilities}. This will not be important in our Fisher matrix analysis with BINGO, as we only need to calculate derivatives around the fiducial model. However, our covariance matrices from \textit{Planck} {were obtained from a MCMC process}, and therefore will depend on these priors. In Fig.~\ref{plot_IDE} we do not distinguish between these regions and plot the results for $Q \propto \rho_{de}$ and $w < -1$.

Our results show that BINGO can put a better constraint on the interaction parameter $\xi_{de}$ than \textit{Planck} for the {model with} $Q \propto \rho_{de}$ and $w > -1$. However, the constraint on the DE EoS is weakened. If $w < -1$, we do not observe appreciable difference in our \hi\,  Fisher matrix  compared with the previous case, but \textit{Planck} possesses better constraints for both the DE EoS and the interaction parameter. In general, the combination with \textit{Planck} yields results of $\delta w \sim 5\%$ and $\sigma_{\xi_{de}} \sim 0.02$ for $Q \propto \rho_{de}$. If the interaction is proportional to the DM energy density, BINGO and \textit{Planck} have opposite behaviors, with BINGO providing tighter constraints on the DE EoS, but allowing a wider uncertainty in the interaction, and \textit{Planck} having looser constraints on the EoS, but tight contours for the interaction parameter. This is expected, as shown in Fig. 1 of \cite{Costa:2019uvk} where the BAO scale is more affected by the interaction at higher redshifts, and was also obtained in Table 10 of \cite{Costa:2016tpb} comparing low-redshift data with {\it Planck} data. Finally, the combination of the two surveys greatly improves the cosmological constraints. We summarize our results in Table \ref{tab:IDE}.

\begin{table}
\footnotesize
\caption{\label{tab:IDE} Expected constraints on the DE EoS and the coupling constant from BINGO, \textit{Planck}, and BINGO + \textit{Planck} combined.}
\begin{center}


\begin{tabular}{|c|S|S|S|}
\cline{2-4}
\multicolumn{1}{c}{} &
\multicolumn{1}{|c|}{BINGO} & 
\multicolumn{1}{|c|}{\textit{Planck}} & 
\multicolumn{1}{|c|}{BINGO + \textit{Planck}} \\
\hline
\multicolumn{1}{|c}{$Q \propto \rho_{de}$, $w > -1$}  &
\multicolumn{1}{c}{} & 
\multicolumn{1}{c}{} & 
\multicolumn{1}{c|}{} 
 \\
 \hline
$\sigma_w$ & 0.37 & 0.072  & 0.047 \\
$\sigma_{\xi_{de}}$ & 0.060 & 0.078 &  0.018 \\
 \hline
 \multicolumn{1}{|c}{$Q \propto \rho_{de}$, $w < -1$}  &
\multicolumn{1}{c}{} & 
\multicolumn{1}{c}{} & 
\multicolumn{1}{c|}{} 
 \\
 \hline
$\sigma_w$ & 0.37 & 0.26 &  0.056 \\
$\sigma_{\xi_{de}}$ & 0.059 & 0.026 & 0.015 \\
 \hline
 \multicolumn{1}{|c}{$Q \propto \rho_{dm}$}  &
\multicolumn{1}{c}{} & 
\multicolumn{1}{c}{} & 
\multicolumn{1}{c|}{} 
 \\
 \hline
$\sigma_w$ & 0.12 & 0.40  & 0.078 \\
$\sigma_{\xi_{dm}}$ & 0.016 & 0.001 &  0.0007 \\
 \hline
\end{tabular}
\end{center}

\end{table}


    

Projected constraints with \hi\,IM experiments for interacting DE models were also considered in \cite{Xu:2017rfo}. Their result for BINGO, however, could be about ten times stronger for the DE EoS and about four times stronger for the interacting parameter under {the model with} $Q \propto \rho_{de}$ and $w > -1$. Although there are some differences in the BINGO setup, we were only able to obtain this level of constraint in combination with {\it Planck} data.

\section{Conclusions}
\label{sec:conclusions}
BINGO will be a single-dish radio telescope designed to observe the LSS using \hi\,  intensity mapping as a tracer of the underlying matter distribution. Therefore, it will provide additional data, susceptible to different systematic effects, that can help improve our current understanding of the late-time cosmological expansion and structure formation. In this work we used the 21cm angular power spectra and the Fisher matrix formalism to forecast the constraining power of BINGO on standard and alternative cosmological models, and analyzed the dependency with different instrument configurations. In summary, we  obtained the following:
\begin{itemize}
    \item If we assume the $\Lambda$CDM model, the fiducial BINGO setup (see Table~\ref{tab:bingo_param}) cannot put competitive constraints on \textit{Planck}. However, their combination can improve the confidence in all cosmological parameters;   the Hubble constant and the DM parameter ($\Omega_ch^2$) are the most significant with an improvement of $\sim 25\%$ in both. This is competitive with current combined constraints from CMB and BAO data.

    \item BINGO plays a more significant role if we leave the DE EoS as a free parameter. In the wCDM model, BINGO alone can establish better constraints than {\it Planck} for the EoS. Their combination can reach $1.1\%$ precision for $H_0$ and $3.3\%$ for $w$ at $68\%$ CL, which have improved by respectively $92\%$ and $87\%$ in comparison with {\it Planck} alone.

    \item Under the CPL parameterization, BINGO + {\it Planck} achieves a $2.9\%$ precision in the Hubble constant, $30\%$ in the DE parameter $w_0$, and $\sigma_{w_a} = 1.2$ for $w_a$ at one standard deviation. Although these constraints can be considered large, BINGO has improved the constraints from {\it Planck} alone by up to $78\%$.

    \item Fixing the CPL parameterization as our fiducial cosmological model, we considered how the cosmological constraints should be affected by several different instrumental scenarios:
    
        \textit{Total observational time}: We considered the impact of the total observational time on our cosmological constraints, which affects the thermal noise level of our experiment. We observe that BINGO will mainly affect the constraints on the Hubble constant and the DE EoS. A five-year experiment can improve these constraints by $\delta\sigma_{H_0} = 25\%$, $\delta\sigma_{w_0}$ and $\delta\sigma_{w_a} \sim 21\%$ compared to the one-year survey. The FoM defined by the error ellipsoid improved by 11 times for BINGO only and 2.7 times in combination with {\it Planck}. Although our constraints have improved for longer total observational time, they significantly flatten after three years, suggesting this as an optimal observational time.

        \textit{Number of feed horns}: The fiducial setup assumes 28 feed horns. Considering the effect on the noise level only, increasing the number of horns {to a maximum value of} 60 can improve the constraints {by at most} $10\%$. Given the costs necessary to increase the number of horns and its connection with the total observational time, it may be more significant to increase the number of horns in such way to cover a larger area in the sky.

        \textit{Number of redshift bins}: The constraints are significantly affected by the number of bins considered; however, we observe small deviations from $N_{\rm{bin}} = 64$ to $N_{\rm{bin}} = 128$, suggesting that we have reached a plateau. As the actual number of redshift bins in the raw data of BINGO will be much larger, this choice must take into account the necessary computational calculations. Our analysis suggests that $N_{\rm{bin}} = 64$ is optimal.

        \textit{Cross-correlations}: Some previous analysis considered the Limber approximation to extract the 21cm information. We analyzed the effect of including {all the spectra} on the final constraints. The importance of cross-correlations increases as we increase the number of bins, but eventually reaches a plateau. For $N_{\rm{bin}} = 32$, near BINGO standard configuration, we obtain improvements of $\delta\textrm{ln}(10^{10}A_s) = 0.4\%$, $\delta{\Omega_bh^2} = 6.8\%$, $\delta{n_s} = 7.2\%$, $\delta{h} = 8.3\%$, $\delta{w_0} = 8.8\%$, $\delta{w_a} = 11\%$, $\delta{\Omega_ch^2} = 22\%$, and $\delta{b_{\textrm{\hi\,}}} = 31\%$.

        \textit{RSD}: RSD can break the degeneracy between $A_s$ and $b_{\textrm{\hi\,} }$ and improve the constraints on the other cosmological parameters. At $N_{\rm{bin}} = 128$ the improvements from RSD are given by $\delta{h} = 1\%$, $\delta{\mathrm{ln}(10^{10}A_s)} = 2.2\%$, $\delta{\Omega_bh^2} = 7.8\%$, $\delta{n_s} = 9.5\%$, $\delta{w_0} = 14\%$, $\delta{w_a} = 34\%$, $\delta{\Omega_ch^2} = 44\%$, and $\delta{b_{\textrm{\hi\,}}} = 176\%$, combining BINGO with \textit{Planck} data.



        \textit{Foreground Residuals}: Foreground residuals will both increase the statistical errors and introduce biases in our final cosmological parameter estimation. We take these effects into account assuming that some sort of foreground removal technique has already been accomplished, and model the residual contamination as the sum of Gaussian processes. We obtain that the statistical error bars are not strongly enhanced; the maximum degradation was of $16\%$ for BINGO only and $6\%$ for BINGO + {\it Planck} with the overall scaling factor $\epsilon_{\textrm{FG}} = 1$ (no foreground removal). On the other hand, we found that the bias in our cosmological constraints will be at maximum $1\sigma$ if $\epsilon_{\textrm{FG}} = 10^{-4}$, and can be significantly smaller if we can achieve smaller values for $\epsilon_{\textrm{FG}}$.

    \item We  also compared the BINGO constraints to those from SKA1-MID band 1 and SKA1-MID band 2. Given the larger surveyed area, deeper redshift range, larger number of antennas, and lower system temperature, BINGO cannot compete with them. However, SKA will have more complicated systematic effects and BINGO can be a pathfinder to better understand them.

    \item In addition to our cosmological parameters, BINGO will help us understand the \hi\,  evolution and distribution. This is affected by the cosmological model and the \hi\,  model considered. Combining BINGO and \textit{Planck} data, we obtained a $2.3\%$ precision fixing the \hi\,  density parameter under the CPL parameterization. Our worst scenario is obtained assuming that $\Omega_{\textrm{\hi\,}}$ and $b_{\textrm{\hi\,}}$ vary with redshift as a free parameter over three groups of redshift bins. In this case we have $\sigma_{\Omega_{\textrm{\hi\,}}^i} \sim 8.5\%$ and $\sigma_{b_{\textrm{\hi\,}}^i} \sim 6\%$.
    
    \item Measurements from the \hi\,  power spectrum can break the geometric degeneracy in the parameter space of $\Lambda$CDM + $\sum m_\nu$. In combination with BINGO, we obtained $\sigma_{\sum{m_\nu}} < 0.14 \, \textrm{eV}$ at $95\%$ CL, which is on the same order of current constraints.
    
    \item BINGO can also help constrain alternative cosmological models breaking degeneracies in the parameter space. In particular, under the $B_0$-parameterization of $f(R)$ gravity, we obtained $\sigma_{B_0} = 3.1 \times 10^{-5}$ with BINGO against $\sigma_{B_0} = 5.3 \times 10^{-2}$ from \textit{Planck}. On the other hand, we forecast that BINGO + \textit{Planck} will be able to put constraints of $\sigma_{\xi_{de}} \sim 0.02$ and $\sigma_{\xi_{dm}} = 0.0007$ on the interacting dark energy parameters.
\end{itemize}

We would like to note some limitations and future extensions of the present work:
\begin{itemize}
    \item First, we   considered the full 21cm angular power spectrum. Although this has more information than the BAO data alone, it is also more contaminated by the \hi\,  physics.

    \item In our analysis we   used the Fisher matrix formalism. It is well known that the Fisher matrix produces the optimal scenario, thus we should expect deviations from the present constraints in a more robust MCMC sampling of the parameter space through the 21cm likelihood. Because of the tomographic nature of our 21cm angular spectra, spanning a 3D volume, the number of necessary computations can increase dramatically  compared to the CMB angular spectra. Therefore, given our computational resources at the moment, we decided to adopt the Fisher matrix formalism, which has been widely used in the literature to forecast cosmological constraints and can provide consistent results.

    \item As discussed before, BINGO will be mainly affected by the presence of foregrounds. Here the bulk of our analysis  considers a perfect foreground removal technique. Of course, this will not be the case, and foreground residuals must be taken into account. In the BINGO companion papers \citep{2020_sky_simulation} and \citep{2020_component_separation}, the foreground cleaning process is further discussed.

    \item The real situation will also be contaminated by other effects. The $1/f$ noise will be the most prominent of them for the BINGO configuration. Other effects that must be taken into account in the final analysis include standing waves, side lobes, RFI, and atmospheric effects.
    
    \item Although BINGO's beam resolution will suppress most nonlinear effects, very thin redshift bins can have a significant contribution from nonlinearities. Nonlinear effects are further discussed in our BINGO companion paper \citep{2020_mock_simulations} and their impact on the cosmological parameters will be analyzed in a future work.
\end{itemize}
\begin{acknowledgements}

The BINGO project is supported by FAPESP grant 2014/07885-0; the support from CNPq is also gratefully acknowledged (E.A.). A.A.C. acknowledges financial support from the China Postdoctoral Science Foundation, grant number 2020M671611. R.G.L. thanks CAPES (process 88881.162206/2017-01) and the Alexander von Humboldt Foundation for the financial support. C.P.N. would like to thank S{\~a}o Paulo Research Foundation (FAPESP), grant 2019/06040-0, for financial support. F.B.A. acknowledges the UKRI-FAPESP grant 2019/05687-0, and FAPESP and USP for Visiting Professor Fellowships where this work has been developed. B.W. and A.A.C. were also supported by the key project of NNSFC under grant 11835009. C.A.W. acknowledges a CNPq grant 2014/313.597. T.V. acknowledges CNPq Grant 308876/2014-8. K.S.F.F. would like to thank FAPESP for financial support grant 2017/21570-0. A.R.Q., F.A.B., L.B. and M.V.S. acknowledge PRONEX/CNPq/FAPESQ-PB (Grant no. 165/2018). V.L. would like to thank S{\~a}o Paulo Research Foundation (FAPESP), grant 2018/02026-0, for financial support. L.S. is supported by the National Key R\&D Program of China (2020YFC2201600). J.Z. was supported by IBS under the project code, IBS-R018-D1. We thank an anonymous referee for his/her very insightful comments.
      
\end{acknowledgements}

\bibliographystyle{aa} 
\bibliography{references.bib} 

\begin{appendix}
\section{Code comparison}
\label{sec:appendix}
In order to check the outputs of the code used throughout this work, we show here a comparison between the {$C_\ell$ values} (and their derivatives) calculated using this code  and those obtained from the \textit{Unified Cosmological Library for
$C_\ell$s} code \citep[\texttt{UCLCL};][]{mcleod2017joint,Loureiro:2018qva}, matching cosmologies as closely as possible. 
\texttt{UCLCL} uses the power spectra and transfer functions from the \texttt{CLASS} code \citep{blas2011cosmic} to construct the angular power spectrum $C_\ell$,
\begin{equation}
    C_\ell^{ij}=\frac{2}{\pi}\int W_{\ell}^i(k)W^j_{\ell}(k) k^2 P(k)dk\,,
\end{equation}
where the indices $i$ and $j$ denote the different redshift bins, $P(k)$ is the underlying matter density field power spectrum at zero redshift, and $W_{\ell}(k)$ is the window function that accounts for projection effects and all the processes involved in the evolution, including RSD \citep[see discussion in][]{Loureiro:2018qva}. 

We set the cosmological parameters to the most recent Planck results  for $\Lambda$CDM as fiducial values \citep[$\Omega_bh^2=0.0224$, $\Omega_ch^2= 0.120$, $h=0.673$, $\Omega_kh^2=0$, $w_0=-1$, $w_a=0$,  $n_s=0.965$, $\ln (10^{10} A_s)=3.096$;][]{Aghanim:2018eyx}. 

\begin{figure}[h]
\centering
\includegraphics[width=1\columnwidth]{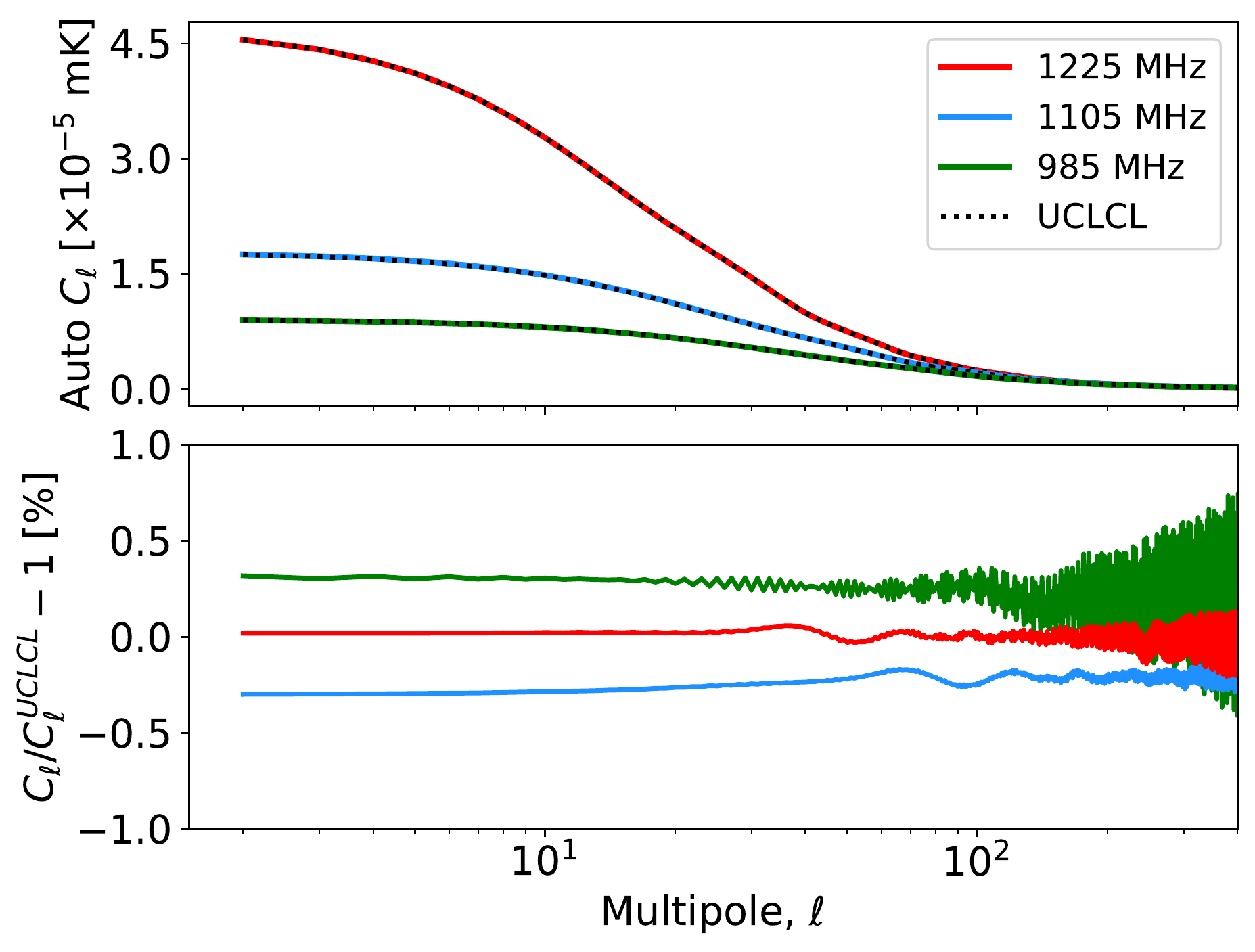}
\caption{Comparison between the auto-correlation $C_\ell^{ij}$s (for $i = j$) calculated in this work and those obtained with the {\tt UCLCL} code at three BINGO redshift bins, centered at the frequencies 985, 1105, and 1225 MHz (redshifts $z \approx 0.44, 0.28$, and $0.16$, respectively), and bandwidths of 10MHz. 
The upper panel shows the {\tt UCLCL} results overplotted as dotted lines on each of the calculated curves (colored lines), while the lower panel shows the percentage relative difference among them. }
\label{fig:auto_cls_frequencies}
\end{figure}

\begin{figure}[h]
\centering
\includegraphics[width=1\columnwidth]{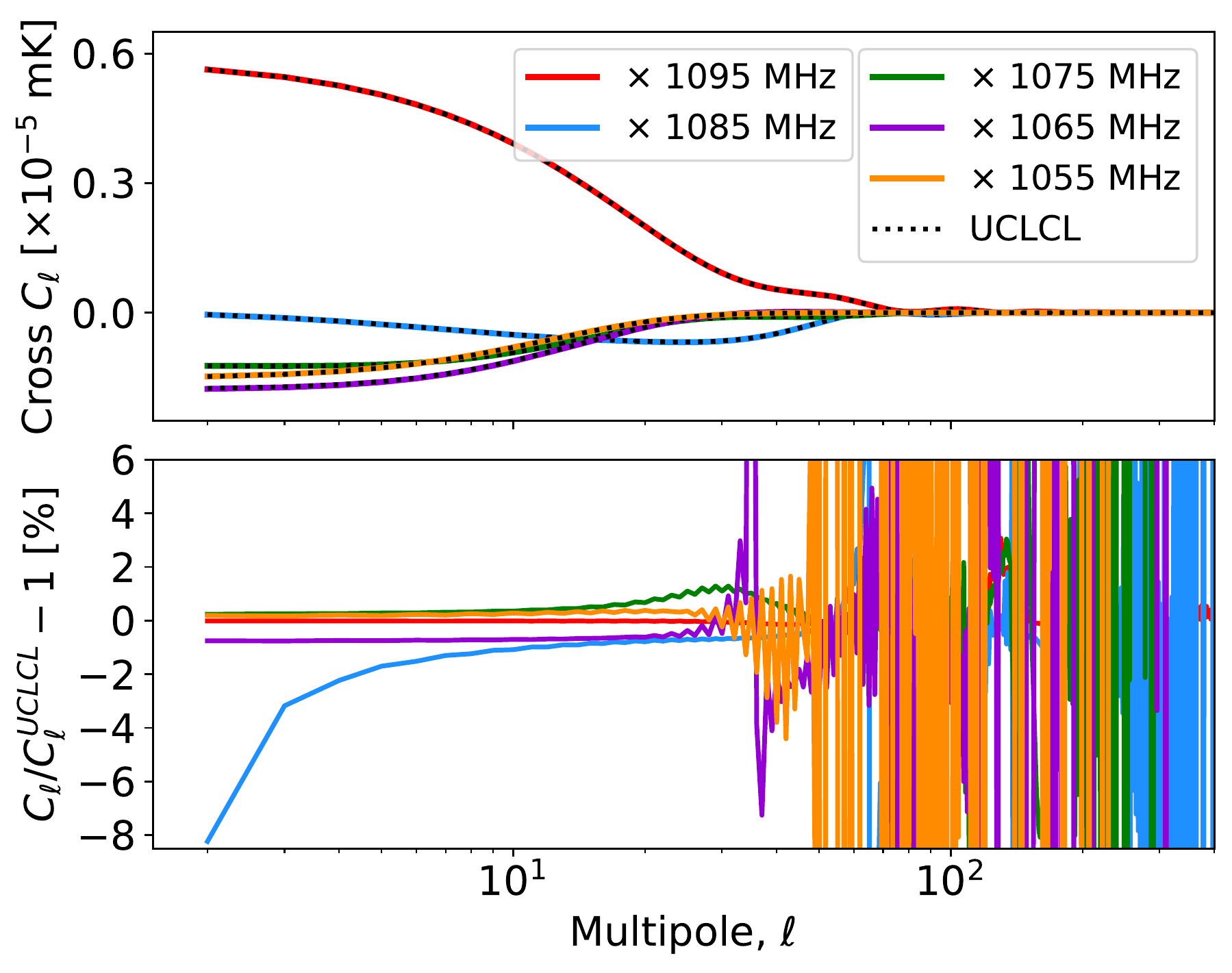}
\caption{Same as Fig. \ref{fig:auto_cls_frequencies}, but for the cross-correlation $C_\ell^{ij}$s ($i \neq j$) between the frequency bin centered at 1105 MHz ($z=0.28$) and the five following frequency bins, with bandwidth of 10MHz. }
\label{fig:cross_cls_frequencies}
\end{figure}

We   performed several tests and comparisons, and here we  present a set of them as illustrative examples in order to give the overall picture of the level of agreement among the results of each code. 
We first compared the results obtained for a total of 30 frequency bands, each with 10 MHz bandwidths, covering the BINGO frequency range. Some examples of the resulting auto- and cross-correlation spectra are shown in the upper panel of Figs.  \ref{fig:auto_cls_frequencies} and \ref{fig:cross_cls_frequencies}, respectively. The lower panel in each figure presents the respective relative difference, in percentage, and shows that the codes deviate from each other by less than 1\% for the auto-$C_\ell^{ij}$s and less than 2\% for the cross-$C_\ell^{ij}$s at multipoles up to $\ell\sim 100$. A larger discrepancy for higher multipoles is introduced not only by small numerical errors (characterized by the noisy behavior), but mainly because the $C_\ell^{ij}$s have very small absolute values at this region, reaching zero, especially in the case of cross-correlation. This is evident from the blue line in Fig. \ref{fig:cross_cls_frequencies}, representing the cross-correlation among the frequency bins centered at 1105 MHz and 1085 MHz, which present $C_\ell^{ij}$ values near  zero at the first multipoles (upper panel), resulting in a larger relative difference at these multipoles as well (lower panel).

\begin{figure}[h]
\centering
\includegraphics[width=1\columnwidth]{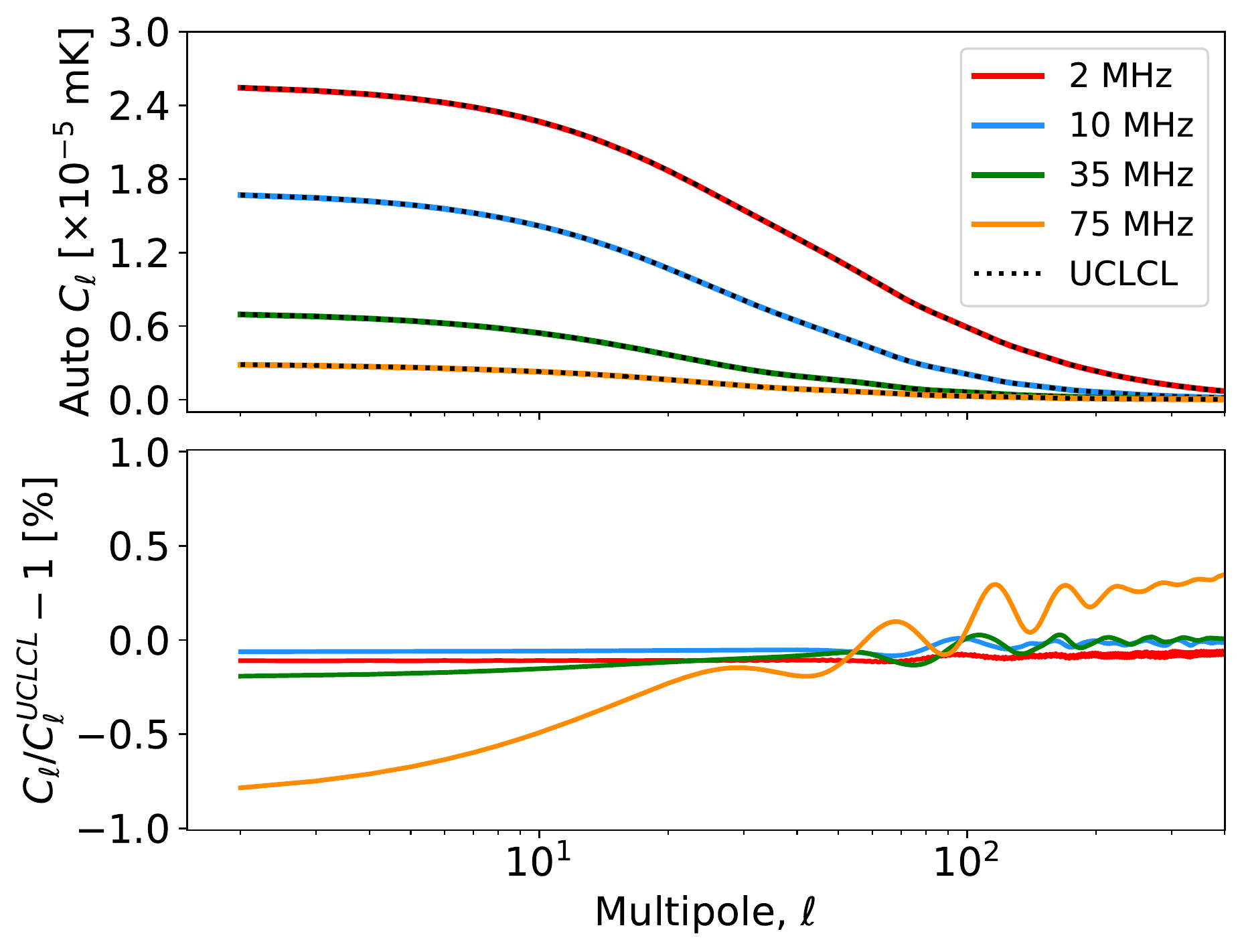}
\caption{Comparison between the auto-correlation $C_\ell^{ij}$s calculated in this work and using the {\tt UCLCL} code for four different bandwidths between 2 and 75MHz, all of them centered at a frequency of 1110MHz. 
Again, the upper panel shows the results overplotted and the lower panel shows the percentage difference among them.}
\label{fig:auto_different_bandwidths}
\end{figure}

We also investigated the influence of different bandwidths when comparing the codes. 
We tested four bandwidths (2 MHz, 10 MHz, 35 MHz and 75 MHz) centered at the frequency of 1110 MHz.
The comparison among some of the auto-$C_\ell^{ij}$s resulting from each code is depicted in Fig. \ref{fig:auto_different_bandwidths}.  
Again, the level of agreement is better than 99.5\% for the thinner bandwidths, while it reaches 99\% for the thicker bandwidth, showing that the smaller the bandwidth the better the level of agreement between the codes. 

\begin{figure}[h]
\centering
\includegraphics[width=1\columnwidth]{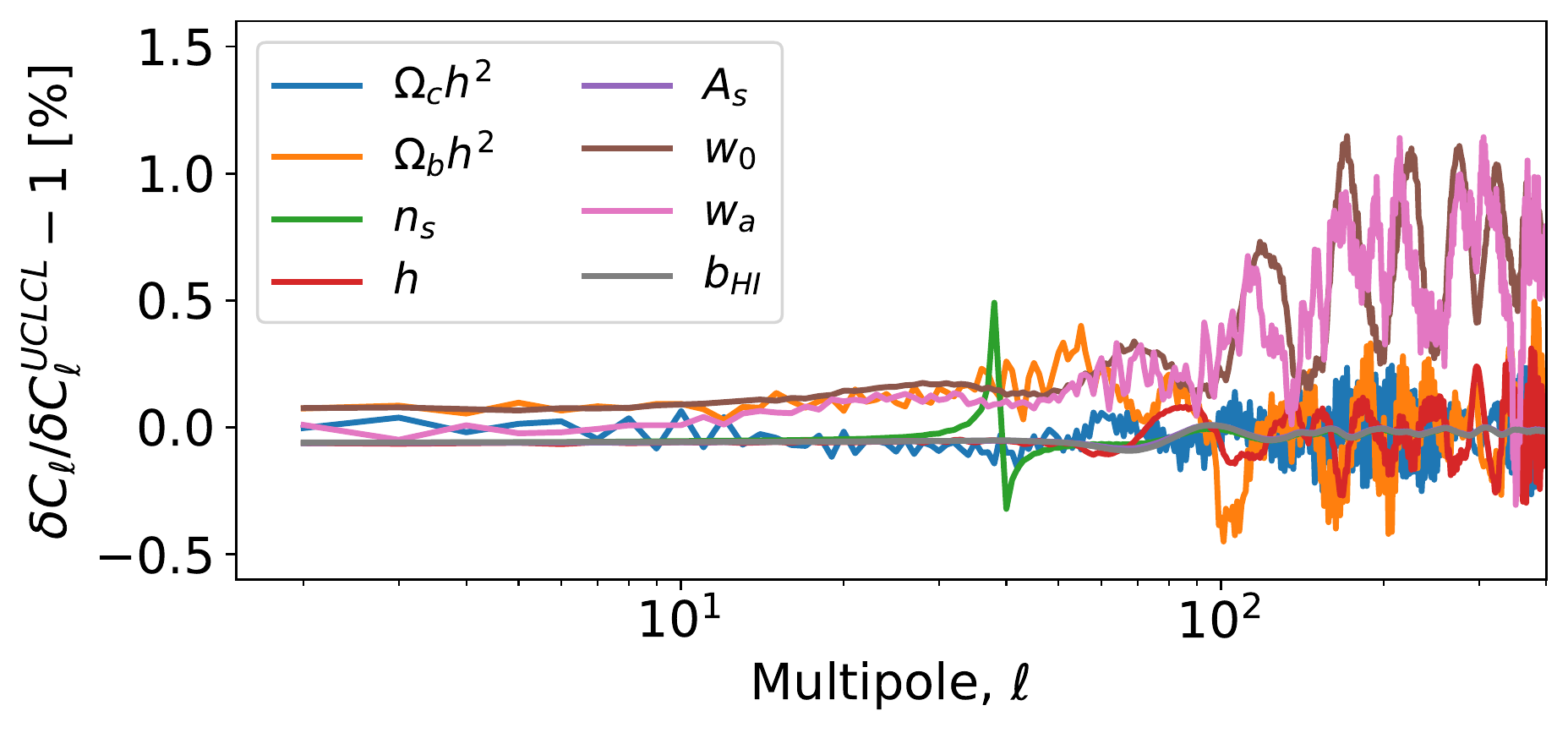}
\caption{Percentage relative difference comparing the derivatives, $\delta C_\ell^{ij} = d C_\ell^{ij} / d x$ (for auto-correlation, $i=j$), with $x$ representing each cosmological and 21cm parameter, calculated in this work and by using the {\tt UCLCL} code. 
All the cases correspond to a frequency channel with a bandwidth of 10MHz centered at 1110MHz.}
\label{fig:derivatives}
\end{figure}

Finally, we compared the two codes in the context of the derivatives of the $C_\ell^{ij}$s with respect to a set of cosmological and 21cm parameters, calculated with a variation of 1\% on the corresponding fiducial values. The comparison, presented in Fig. \ref{fig:derivatives} for a bandwidth of 10 MHz centered at 1110 MHz, shows an agreement of more than 99\%, for all parameters considered, for multipoles up to $\ell \sim 100$, and of more than 98\% for larger multipoles. The same comparison was performed for  other different bandwidths, obtaining results similar to  those observed in Fig. \ref{fig:derivatives} for bandwidths of 2 MHz and 35 MHz, and a larger discrepancy for the broader band, 75 MHz, but still following the expected from Fig. \ref{fig:auto_different_bandwidths}.

\end{appendix}

\end{document}